\documentclass[aps,prl,preprint,groupedaddress]{revtex4-1}
 \usepackage{graphicx}
 \usepackage{dcolumn}
 
 \usepackage{bm}
 \usepackage{amsmath}
 \usepackage{color}
 \usepackage{amssymb}
 \usepackage{mathrsfs}
\usepackage{MnSymbol}%
\usepackage{wasysym}%
\usepackage{gensymb}
\usepackage{framed,color}
\usepackage{enumitem}
\definecolor{shadecolor}{rgb}{1,0.9,0.3}
\usepackage{times}
\usepackage[normalem]{ulem}
\usepackage{lineno}

\usepackage{textcomp}
\usepackage{calc}
\usepackage{slashed}
\usepackage{amsthm,amsmath,amsfonts,amssymb,verbatim,color}
\usepackage[T1]{fontenc}
\usepackage[colorlinks=true,citecolor=blue,linkcolor=blue,urlcolor=blue]{hyperref}

\begin{document}

\title{Fractional Little-Parks effect observed in a topological superconductor}

\author{Yufan Li, Xiaoying Xu, Shu-Ping Lee, and C. L. Chien}

\bigskip

\affiliation{Department of Physics and Astronomy, The Johns Hopkins University, Baltimore, MD 21218, USA}

\date{\today}

\begin{abstract}
	In superconductors, the condensation of Cooper pairs gives rise to fluxoid quantization in discrete units of $\Phi_0 = hc / 2e$.
	The denominator of $2e$ is the signature of electron pairing, which is evidenced by a number of macroscopic quantum phenomena, such as the Little-Parks effect and the Josephson effect, where the critical temperature or the critical current oscillates in the period of $\Phi_0$.
	Here we report the observation of fractional Little-Parks effect in mesoscopic rings of epitaxial $\beta$-Bi$_2$Pd, a topological superconductor.
	Besides $\Phi_0$, novel Little-Parks oscillation periodicities of $2\Phi_0$, $3\Phi_0$ and $4\Phi_0$ are also observed, implying quasiparticles with effective charges being a fraction of a Cooper pair. 
	We show that the fractional Little-Parks effect may be closely related to the fractional Josephson effect, which is a key signature of chiral Majorana edge states.
\end{abstract}

\maketitle

\begin{text}
The most distinctive feature of superconductivity is the presence of the complex-valued many-particle wave function that sustains phase coherence over macroscopic distances. 
One of its experimental manifestations is the fluxoid quantization in a superconducting ring.
The single-value nature of the wave function dictates a universal phase change of $2\pi$ for any closed path around the ring.
As a result, the fluxoid can only take on quantized values in integer steps of $\Phi_0 = hc / e^*$, where $h$ is the Planck constant, $c$ is the speed of light, and $e^*$ the effective charge.
As a key evidence of electron pairing, $e^*$ is found to be $2e$ of the Cooper pairs \cite{Byers1961}.
The quantization unit of $\Phi_0$, or $2\pi$-periodicity, has been confirmed by numerous experiments including magnetometry measurements \cite{Deaver1961,Doll1961}, the Little-Parks effect \cite{Little1962} and the Josephson effect~/~quantum interference experiments \cite{Rowell1963,Jaklevic1964}.

Different quantization periodicities other than $\Phi_0$ have also been proposed.
Quantization period of $hc/e$, or $4\pi$-periodicity, has been previously suggested for unconventional superconductors (SCs) \cite{Juricic2008,Loder2007,Zhu2010}, and small-size $s$-wave superconducting rings whose dimensions become comparable to the coherence length \cite{Vakaryuk2008,Wei2008,Loder2008}.
Despite the theoretical consensus, experimental confirmation of such effects remain elusive \cite{Sochnikov2010}.
On the other hand, $4\pi$-periodicity has been recently proposed for topological superconductors (TSCs) as a signature of Majorana fermions \cite{Kitaev_2001,Kwon2003,fu_josephson_2009,lutchyn_majorana_2010,lee_electrical_2012}.
Many experiments have been embarked to search for the effect, generally focusing on proximity-induced superconductivity, such as junctions of conventional $s$-wave SCs and strong spin-orbit coupling materials.
The most intriguing experimental indications so far are the missing odd Shapiro steps \cite{wiedenmann_4-periodic_2016,Bocquillon2016,Yu2018,li_4-periodic_2018,wang_$4ensuremathpi$-periodic_2018} and halving of the radiation frequency of AC Josephson effect \cite{Deacon2017,Laroche2019}.
More exotic $8\pi$-periodicity has also been proposed in several different theoretical settings \cite{zhang_time-reversal-invariant_2014,Peng2016a,Malciu2019}, but awaiting experimental verifications.

Here we report multiple novel quantization periodicities observed in the Little-Parks effect of epitaxially-grown $\beta$-Bi$_2$Pd thin films, an intrinsic topological superconductor \cite{Sakano2015,Iwaya2017,Lv2017,Li2019}.
Besides the conventional $2\pi$-periodicity, we also found the presence of $4\pi$-periodicity consistent with single electron tunneling that might originate from Majorana modes. 
Novel $6\pi$- and $8\pi$-periodicities are also observed. 
The non-$2\pi$-periodicities imply the effective charge $e^*$ being a fraction of the electric charge of a Cooper pair. 
The origin of such fractional Little-Parks effect deserves further theoretical deliberation.
We explore the relation between the fractional Little-Parks effect and the fractional Josephson effect \cite{Kitaev_2001,Kwon2003,fu_josephson_2009,lutchyn_majorana_2010,lee_electrical_2012,zhang_time-reversal-invariant_2014,Peng2016a,Malciu2019}. 
An analytical solution of a 1D-Kitaev chain is presented, which reveals the $4\pi$-periodicity of the Little-Parks oscillation. 

Earlier experiments have reported the observation of spin-polarized topological surface state in $\beta$-Bi$_2$Pd \cite{Sakano2015,Iwaya2017}, and the zero-bias conductance in the vortex cores \cite{Lv2017}.
Very recently, we have observed half-quantum fluxoid in polycrystalline rings of $\beta$-Bi$_2$Pd \cite{Li2019}.
This indicates the pairing state of $\beta$-Bi$_2$Pd must be anisotropic, suggesting spin-triplet pairing, consistent with the expectation of a TSC.
In this work, we prepare 70~nm-thick epitaxial $\beta$-Bi$_2$Pd/SrTiO$_3$(001) and $\beta$-Bi$_2$Pd/MgO(001) thin films by magnetron sputtering, with details in the Supplementary Information \cite{SM}.
The thin films are subsequently patterned into mesoscopic ring devices by electron-beam lithography.
We have prepared multiple designs of the epitaxial ring structures with various ring sizes and geometries that conclusively demonstrated the fractional Little-Parks effect.

The Little-Parks effect concerns the quantum oscillation of the superconducting critical temperature, $T_c$, as a function of the applied magnetic flux threading through the enclosed area of a superconducting ring, reflecting the periodic oscillation of the free energy \cite{Little1962}.
The $T_c$ oscillation is experimentally observed in the electrical resistance of the ring at temperatures just below $T_c$. 
The Little-Parks oscillation with integer fluxoid quantization of $n\Phi_0$, where $n$ is an integer number, has been observed in nearly all categories of SCs, including $s$-wave SCs \cite{Parks1964,vaitiekenas_flux-induced_2018}, $d$-wave SCs \cite{Gammel1990,Sochnikov2010}, Sr$_2$RuO$_4$ \cite{cai_unconventional_2013}, in \textit{both} single-crystalline and polycrystalline materials.
On the other hand, fluxoid quantization may not necessarily be the \textit{only} mechanism for the free energy oscillation.
For example, it has been proposed that the chiral Majorana edge state may introduce $4\pi$-Josephson effect in a long Josephson junction comprising topological $p+ip$ superconductors, manifesting as a $2\Phi_0$-periodicity modulation to the oscillation of $I_c$ \cite{lee_electrical_2012}.
The oscillation of the critical point reflects the same $4\pi$-periodic oscillation also occurring to the free energy.
One may therefore expect the Little-Parks effect of a TSC to display a $4\pi$-periodic oscillation of $T_c$.

As we have earlier reported in polycrystalline (001)-textured $\beta$-Bi$_2$Pd thin films, the Little-Parks oscillation can exhibit half-quantum fluxoids of $(n+\frac{1}{2})\Phi_0$ as a result of anisotropic pairing state \cite{Li2019}.
For triplet SCs, the gap has odd parity and experiences a sign change upon momentum inversion. 
The sign change can facilitate a $\pi$ phase shift at crystalline grain boundaries as first predicted theoretically \cite{geshkenbein_vortices_1987}. 
Such $\pi$ junctions do not exist in single crystals due to the categorical absence of grain boundaries, therefore half-quantum fluxoid is not expected in epitaxial thin film samples. 
Indeed, no half-quantum fluxoid can be observed in epitaxial rings of $\beta$-Bi$_2$Pd.  
In Figs.~1b and 1c we present side-by-side the results of Little-Parks effect obtained from polycrystalline and epitaxially-grown thin films.
Both samples are fabricated to the same device design and share nearly the same geometrical dimensions. 
While the $\pi$ phase shift is present in the polycrystalline sample, it is \textit{completely absent} in the epitaxial ring device of $\beta$-Bi$_2$Pd/SrTiO$_3$ (Device A). 
Indeed, with no crystalline grain boundaries, \textit{none} of the 30 rings of epitaxial thin films exhibits a $\pi$-phase shift with a half-quantum flux, as we expected. 

On the other hand, the Little-Parks effect in the epitaxial $\beta$-Bi$_2$Pd ring manifests striking new features. 
While the oscillation no longer shows the $\pi$ phase shift, the period of the oscillation is not simply $\Phi_0$, but evidently departs from the $2\pi$-periodicity. 
As shown in Fig. 1d, after a smooth background has been subtracted from the raw data presented in Fig.~1c, the Little-Parks oscillation shows not $2\pi$-periodicity of $\Phi_0$, but predominately $4\pi$-periodicity of $2\Phi_0$. 
The emergence of this new $4\pi$-periodicity may be better examined by the Fourier transform analysis of the Little-Parks oscillations, as shown in Fig.~1e, where the spectral peak occurs at the frequency corresponding to $2\Phi_0$, or the $4\pi$-periodicity. 
Being dwarfed by the overwhelming $4\pi$-periodicity, the presence of the conventional $2\pi$-periodicity is only hinted by the kink-like feature near $\pm~3~\Phi_0$ in Fig.~1c. 
In fact, a small $2\pi$-periodic component can be revealed by analyzing the Fourier transform spectra, detailed in the Supplementary Information \cite{SM}.

We present another example where both $2\pi$- and $4\pi$-periodicities are unambiguously visible from the Little-Parks oscillation.
Fig.~2a shows the temperature dependence of the Little-Parks effect in Device B, a 450 nm-size ring fabricated using an epitaxial $\beta$-Bi$_2$Pd/MgO thin film.
Prominent $2\pi$-periodic oscillation is accompanied by a $4\pi$-periodic component for temperatures of 3.0~K and 3.05~K.
Above this temperature range, the $4\pi$-periodicity becomes overwhelming in contrast to the diminishing $2\pi$-component.
Below 3.0~K, more interestingly, a $6\pi$-periodic component appears to emerge at the low field range for $\Phi < 6\Phi_0$, before giving way to the $4\pi$-periodicity at higher fields for  $\Phi > 6\Phi_0$.
The complex temperature-dependent evolutions of various periodicities may be best demonstrated in the Fourier transform spectra as shown in Fig.~2b.
The amplitude of the conventional $2\pi$-periodicity is more prominent at around 3.0~K, but abating in magnitude towards both lower and higher temperatures.
A strong $6\pi$-periodic peak appears at the low temperatures, while the $4\pi$-periodic component intensifies with ascending temperatures and dominates at 3.1~K. 

The unconventional periodicities are not the results of spurious Fourier analyses. 
Fourier analyses of the Little-Parks oscillations from polycrystalline $\beta$-Bi$_2$Pd ring with $\pi$ phase shift (Fig.~S4a) and Nb ring with conventional integer flux quantization (Fig.~S4b) both reveal only $2\pi$-periodicity with no detectable contributions of other non-$2\pi$-periodicities. 
The non-$2\pi$-periodicities are observed only in epitaxial $\beta$-Bi$_2$Pd rings.

We present yet another example where the Little-Parks oscillation is exclusively dominated by the $6\pi$-periodicity.
Fig.~3 shows the results obtained from Device C, a 900 nm-size ring of epitaxial $\beta$-Bi$_2$Pd/SrTiO$_3$.
The $6\pi$-periodicity is unequivocally demonstrated in both the resistance oscillation (Fig.~3a) and the Fourier transform spectra (Fig.~3b).
The persistence of the $6\pi$-periodicity shows little temperature dependence.
Yet an additional high-pitch oscillation with the period of $\frac{1}{2}\Phi_0$ emerges at the low temperatures, most prominently at 2.5~K (Figs.~3a and S9b).
This $\pi$-periodicity, whose peak is not immediately recognizable from the Fourier transform spectra presented in Fig.~3b due to small oscillation amplitude, can be unambiguously revealed by further analyzing the spectrum (Fig.~S9a) \cite{SM}.
It has been proposed that the half-quantum vortices may be present as a result of spin-triplet pairing \cite{volovik_line_1976,Salomaa1985,chung_stability_2007} and was sought after in $^3$He-$A$ \cite{autti_observation_2016} and Sr$_2$RuO$_4$ \cite{jang_observation_2011,yasui_little-parks_2017}.
In our case, we have observed the $\frac{1}{2}\Phi_0$-periodicity in about one tenth of the devices \cite{SM}.
Its occurrence is far less frequent than the other periodicities of $4\pi$, $6\pi$, and $8\pi$, which is to be discussed next.

The Little-Parks oscillation of $\beta$-Bi$_2$Pd may also manifest $8\pi$-periodicity.
An example is shown in Fig.~3c as the result obtained from Device D, a 600 nm-size ring of epitaxial $\beta$-Bi$_2$Pd/SrTiO$_3$.
It is clear that the oscillation period of $4\Phi_0$ dominates at all temperatures, while a smaller oscillatory component with the ordinary period of $\Phi_0$ is also present.
Prominent peaks corresponding to the $8\pi$-periodicity are evident in the Fourier transform spectra shown in Fig.~3d.
A spike-like feature, centered at the zero magnetic field, emerges for temperatures higher than 2.7~K, as shown in Fig.~3c.
We note that this spike-like feature is not unique to Device D, but also observed in other samples, such as Devices A and C.
Its sharp decay at finite magnetic field and the non-repeating nature indicate that the spike is not part of the Little-Parks oscillation, but more likely a signature of weak localization \cite{SM}.

The fluxoid quantization, which produces $2\pi$-periodicity, is universal for a superconducting loop, regardless of the details of the particular device, such as size, shape, presence of defects, \textit{etc}.
In contrast, the novel non-$2\pi$-periodicities are only observed in epitaxial $\beta$-Bi$_2$Pd, with noticeable variations among devices. 
The variation is not determined by the ring geometry, as devices with nominally identical shapes may exhibit different combination of periodicities \cite{SM}. 
An individual device often manifests a superposition of multiple periodicities. 
For a particular epitaxial ring device, although one cannot predict the predominate periodicity \textit{a priori}, it is certain that at least one of these non-$2\pi$-periodicities shall be present. 
In Fig.~4 we display the number of occurrences of different periodicities from a total of 30 epitaxial $\beta$-Bi$_2$Pd devices. 
While the conventional $2\pi$-periodicity is still the most commonly observed, the occurrence of the $4\pi$-, $6\pi$- and $8\pi$-periodicities are only slightly less. 
In the Supplementary Information we present more examples for each non-$2\pi$-periodicities \cite{SM}.

With the exception of the $\pi$-periodicity \cite{yasui_little-parks_2017}, non-$2\pi$-periodicities are not expected for the Little-Parks effect. 
They indicate $e^*$ being $e$, $\frac{2}{3}e$ and $\frac{1}{2}e$ respectively instead of $2e$; concluding the presence of any individual one in a superconductor is potentially highly significant. 
First and foremost, concluding non-$2\pi$-periodicities is contingent on the correct determination of the conventional $2\pi$-periodicity, which can be affected by the fidelity of nanofabrication and adversary effects such as flux focusing.
We note that in many of the devices the $2\pi$-periodicity coexists with other periodicities (e.g. Devices B and D), which allows determining $\phi_0$ in a self-consistent manner.
In general we find the observed $\phi_0$ periods in good agreement with the expected values calculated from the design geometry.
A summary is presented in Table~S1 \cite{SM}.
As control experiments, we have examined the Little-Parks effect in polycrystalline rings of Nb/Si, $\beta$-Bi$_2$Pd/Si \cite{Li2019}, and $\alpha$-BiPd/SrTiO$_3$ \cite{xu2020spintriplet}, where only the conventional $2\pi$-periodicity is observed.
These studies employed the same device design geometries as those of this work, and the observed $\phi_0$-oscillation periods agree with expected values.
At any rate, the non-$2\pi$-periodicities differ from the conventional $2\pi$-periodicity by at least a factor of 2, too large to be discounted by flux focusing or uncertainties in calculating the effective loop area.
Neither can these potential artifacts explain the coexistence of multiple periodicities, amply on display in Devices B, C and D.

It is also known that anomalies in flux transitions may occur at much lower temperatures when $T \ll T_c$.
The system could stay locked in a metastable state before a sudden collapse into the ground state, resulting a flux jump by an arbitrary integer number of $\phi_0$ \cite{Pedersen2001,Vodolazov2003,Berger2003,Arutyunov2004}.
The onset of such an effect requires minimal thermal fluctuation at very low temperatures, therefore cannot occur in the resistive state in the vicinity of $T_c$, where the Little-Parks effect is observed \cite{Berger2003}.
In these flux jump experiments, often utilizing conventional $s$-wave superconductors, the Little-Parks effect is exploited as a control experiment because it manifests pure $\phi_0$-oscillation \cite{Arutyunov2004}.
The avalanche from the metastable state to the ground state results an asymmetric saw-tooth pattern in the oscillations of the magnetic flux \cite{Pedersen2001,Vodolazov2003} and the critical current \cite{Arutyunov2004}, in contrast to the smooth sinusoidal Little-Parks oscillations observed in this work.
When the flux jumps, the relaxation to the ground state needs to be thorough, thus inconsistent with the observed superposition of multiple periodicities in $\beta$-Bi$_2$Pd.
We conclude that the non-$2\pi$-periodicities do not originate from multiple flux jumps.

We consider a theoretical model of a ring structure of a 1D Kitaev chain with a single Josephson junction. 
It can be shown that the $T_c$ manifests Little-Parks oscillation with the period of $2\Phi_0$ \cite{SM}, as a result of forming Majorana edge states, resembling the $4\pi$-Josephson effect of the same origin \cite{Kitaev_2001,lee_electrical_2012}.
Under this context, it may also help to understand why this effect is not observed at all in polycrystalline specimens.
In that case, counter-propagating Majorona edge states confined by the opposite crystal grain boundaries may tend to cancel each other due to the limited grain sizes on the order of 20~nm \cite{Li2019}.
The Josephson junction modeled in our theoretical analysis may translate to crystalline defects inside the loop of the ring device, which could act as weak links.
We note that the model does not reproduce the $6\pi$- and $8\pi$-periodicities, for the obvious reason that the corresponding mechanisms are not included in the simplified 1D Kitaev chain.
Nevertheless, our theoretical analysis establish that the fractional Josephson effect may manifest itself as $T_c$ oscillation; the effect can be subsequently termed fractional Little-Parks effect.

As concluding remarks, we discuss the implications of our findings.
We have investigated the Little-Parks effect of epitaxial $\beta$-Bi$_2$Pd thin films and discovered non-$2\pi$-periodicities which corresponds to effective charges as fractions of a Cooper pair. 
They appear to suggest quasi-particles with charges of $e$, $\frac{2}{3}e$ and $\frac{1}{2}e$. 
This fractional Little-Parks effect, to our best knowledge, has not been observed in any other superconducting materials.
Even for $\beta$-Bi$_2$Pd, such an effect does not appear in polycrystalline samples.
Future theories to comprehensively understand the effect need to reconcile these facts.

Our proposed model based on Majorana edge states can readily explain the $4\pi$-periodicity (i.e. $2\Phi_0$ oscillation) and account for the differences between epitaxial and polycrystalline samples.
Although only one junction is considered in the model, we note that it would not make qualitative differences if the loop contains multiple such junctions, since they all produce the same periodicity.
One expects a superposition of the $2\pi$-periodicity from the flux quantization, and the $4\pi$-periodicity from Majorana edge states.
The competition among respective free energy terms may give rise to particular temperature- or field-dependences of each periodicity. 

It is often argued that the $4\pi$-periodicity can only be observed as an AC effect but not in the DC limit, due to quasiparticle poisoning, a term that describes the Majorana state relaxing to its ground state by parity switching, which restores the conventional $2\pi$ current-phase relation in a Josephson junction \cite{Kwon2003}. 
There is an implicit presumption that the time scale between the parity switching events is always longer than the ground state relaxation time, which is not necessarily true \cite{Lee2014}. 
If the parity switching processes happen faster than the relaxation to the ground state, the Majorana state can be immune to the quasiparticle poisoning effect and allow the $4\pi$-periodicity to be observed in the DC limit, which could be the case for the fractional Little-Parks effect. 

One may further postulate that different types of junctions may be present which can give rise to $6\pi$- or $8\pi$-periodicities. 
Manifestation of a particular periodicity depends on the details of the defects enclosed in the loop, which accounts for the observed variation of behaviors among ring devices. 
Future experiments should try to obtain monochromatic oscillations by better controlling the defects.
Several mechanisms have been proposed for the $8\pi$-Josephson effect \cite{zhang_time-reversal-invariant_2014,Peng2016a,Malciu2019}.
In principle, an analysis of the free energy, similar to the one we demonstrated here \cite{SM}, may show that the $8\pi$-periodicity would also show up in the Little-Parks effect. 
Possibilities of non-topological mechanisms should also be explored. 
The shot noise in normal metal/superconductor junctions produces the appearance of a $\frac{2}{3}e$ fractional charge, an effect that has no topological origin \cite{Jehl2000}.
Absence of any immediate connection to the Little-Parks effect, the resemblance to the $6\pi$-periodicity reported here seems more coincidental than substantial.
Nevertheless, these issues invite further theoretical investigations.

\textbf{Acknowledgments}

This work was supported by the U.S. Department of Energy, Basic Energy Science, Award Grant No. DESC0009390. The e-beam lithography was conducted at the University of Delaware Nanofabrication Facility (UDNF). We thank Kevin Lister for assistance in the nanofabrication processes.

\end{text}

\newpage

\begin{figure}
	\centering
	\includegraphics[width=14cm]{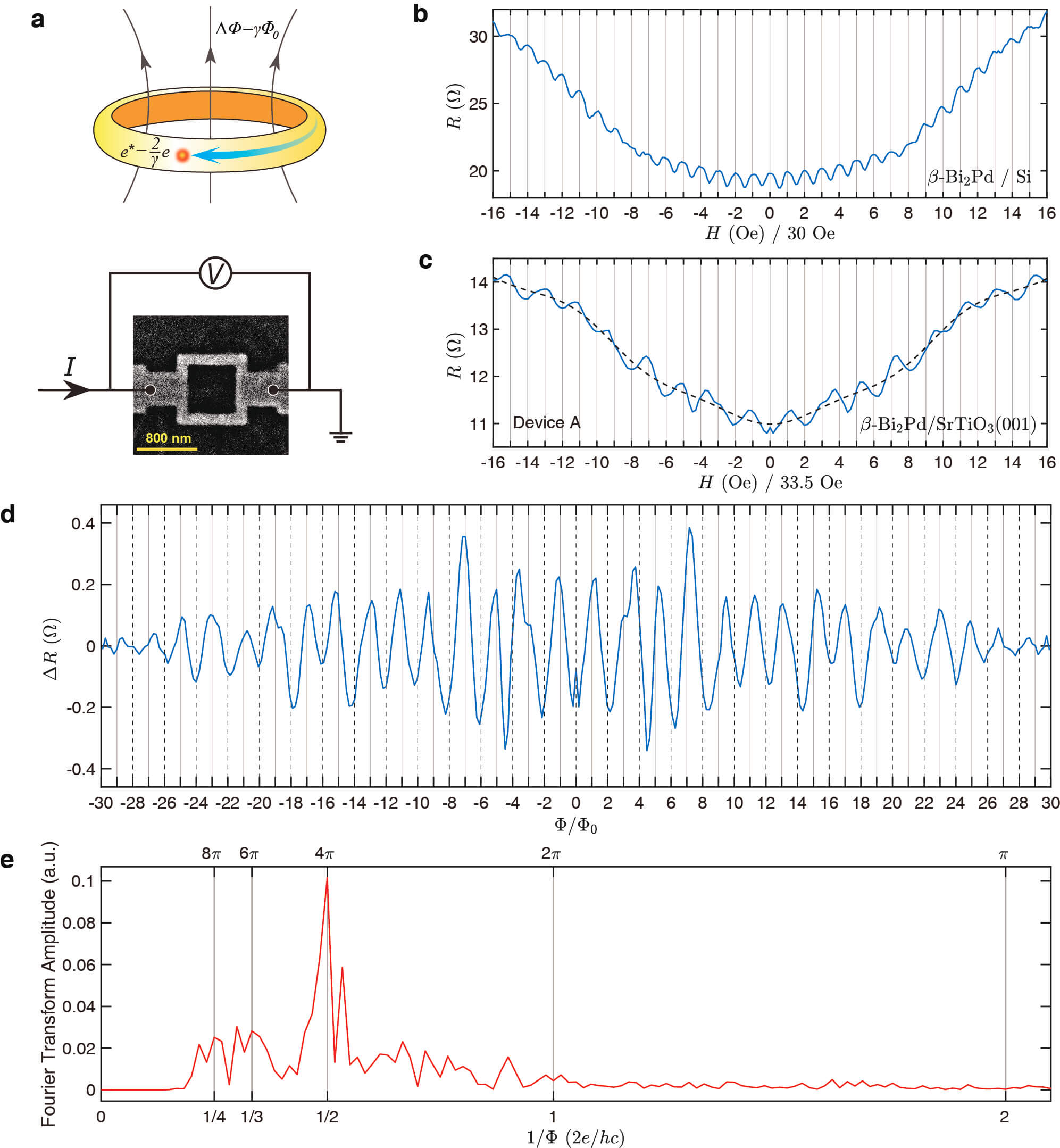}
	\caption{\label{fig1}
		\textbf{Little-Parks effect in polycrystalline and epitaxial $\beta$-Bi$_2$Pd thin film specimens.}
		\textbf{a}, Top: relation between the effective charge and the quantization periodicity. $\gamma$ is an integer number. Bottom: schematic drawing of the experimental setup and the scanning electron microscope image of a representative ring device (800 nm by 800 nm in area).
		\textbf{b}, The Little-Parks effect of a polycrystalline $\beta$-Bi$_2$Pd ring device measured at 2.5~K, reproduced from Ref. \cite{Li2019}. The device geometry is shown in (\textbf{a}). The expected oscillation period of $\Phi_0$ is 32.3~Oe.
		\textbf{c}, The Little-Parks effect of Device A, a ring device made of an epitaxial $\beta$-Bi$_2$Pd/SrTiO$_3$(001) thin film, measured at 2.7~K. The design geometry is identical to the polycrystalline device shown in (\textbf{b}). The black dashed line denotes a smooth background.
		\textbf{d}, The Little-Parks oscillation of Device A at 2.7~K, with the smooth background subtracted from the raw data. $\Phi_0$ corresponds to an oscillation period of 33.5~Oe. 		
		\textbf{e}, The Fourier transform of the Little-Parks oscillation data in (\textbf{d}).
	}
\end{figure}

\clearpage

\begin{figure}
	\centering
	\includegraphics[width=17cm]{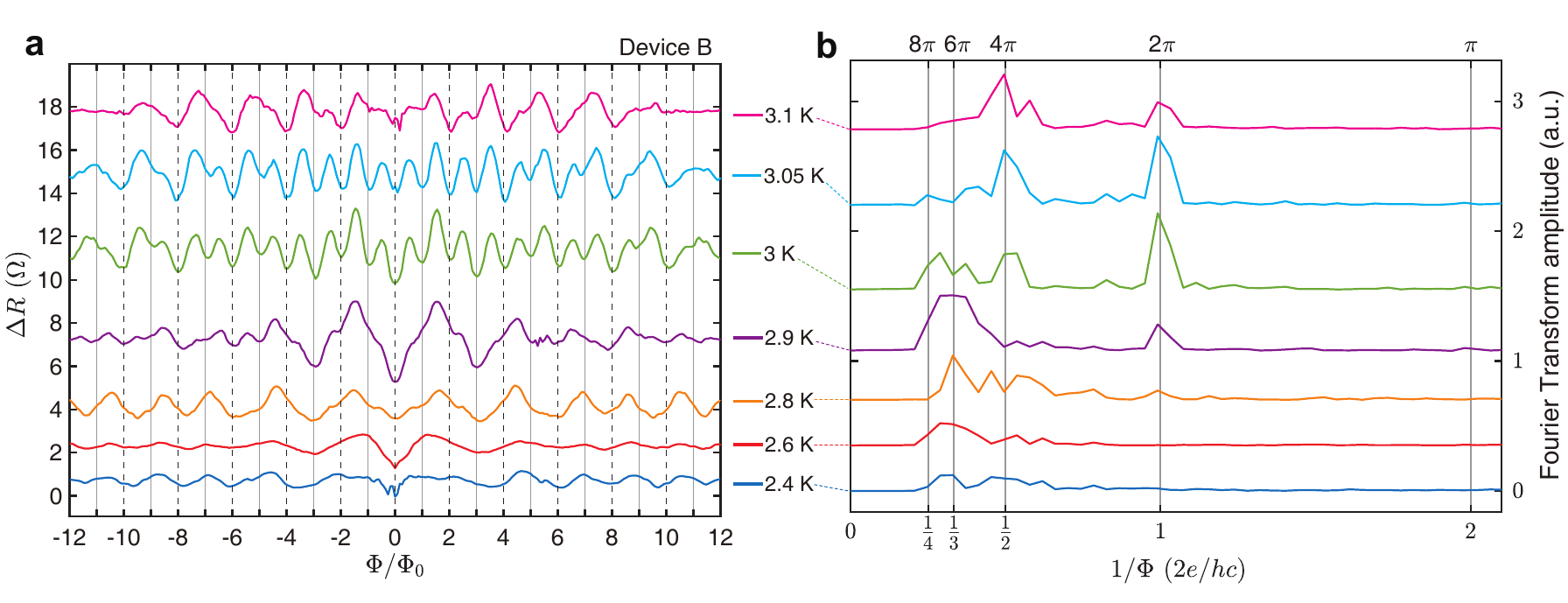}
	\caption{\label{fig2}
		\textbf{Little-Parks effect of Device B and $4\pi$-periodicity.}
		\textbf{a}, The Little-Parks oscillations of Device B, a ring device of epitaxial $\beta$-Bi$_2$Pd/MgO(001), with design geometry of 450~nm by 450~nm in area. The period for $\Phi_0$ expected from the design geometry is 102~Oe, whereas it is determined to be 116~Oe from the experimental data, as shown in the plot.
		\textbf{b}, The Fourier transform of the Little-Parks oscillation data in (\textbf{a}). Both figures share the same color code for various temperatures as shown in the middle of the two figures.
	}
\end{figure}
\clearpage

\begin{figure}
	\centering
	\includegraphics[width=17cm]{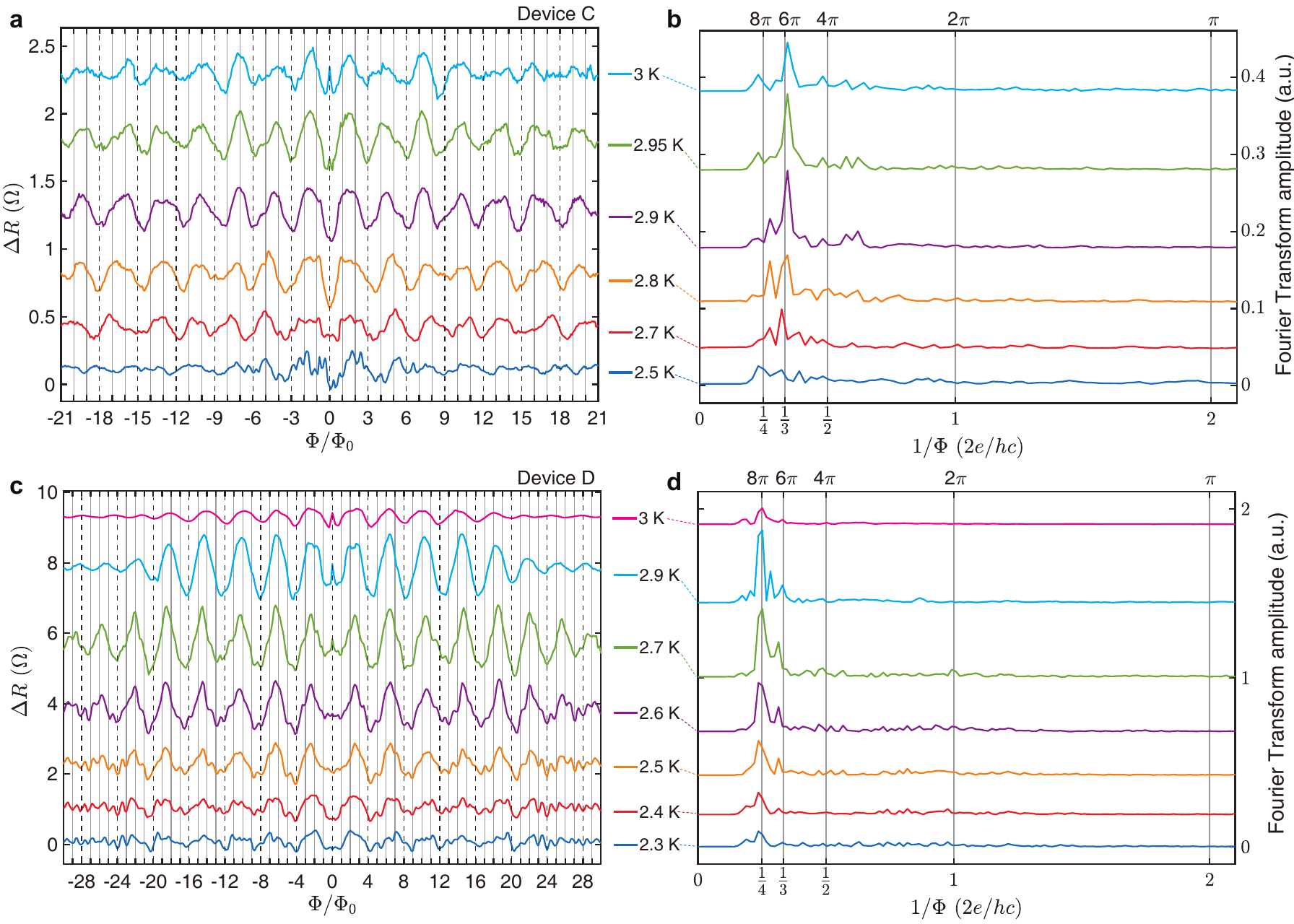}
	\caption{\label{fig3}
		\textbf{$6\pi$- and $8\pi$-periodicities.}
		\textbf{a}, The Little-Parks oscillations of Device C, a ring device of epitaxial $\beta$-Bi$_2$Pd/SrTiO$_3$(001), with design geometry of 900~nm by 900~nm in area. The period for $\Phi_0$ expected from the design geometry is 25.5~Oe, whereas it is determined to be 27.6~Oe from the experimental data, as shown in the plot.
		\textbf{b}, The Fourier transform of the Little-Parks oscillation data in (\textbf{a}). Both (\textbf{a}) and (\textbf{b}) share the same color code for various temperatures as shown in the middle of the two figures.
		\textbf{c}, The Little-Parks oscillations of Device D, a ring device of epitaxial $\beta$-Bi$_2$Pd/SrTiO$_3$(001), with design geometry of 600~nm by 600~nm in area. The period for $\Phi_0$ expected from the design geometry is 57.4~Oe, whereas it is determined to be 60.0~Oe from the experimental data, as shown in the plot.
		\textbf{d}, The Fourier transform of the Little-Parks oscillation data in (\textbf{c}). Both (\textbf{c}) and (\textbf{d}) share the same color code for various temperatures as shown in the middle of the two figures.
	}
\end{figure}
\clearpage

\begin{figure}
	\centering
	\includegraphics[width=10cm]{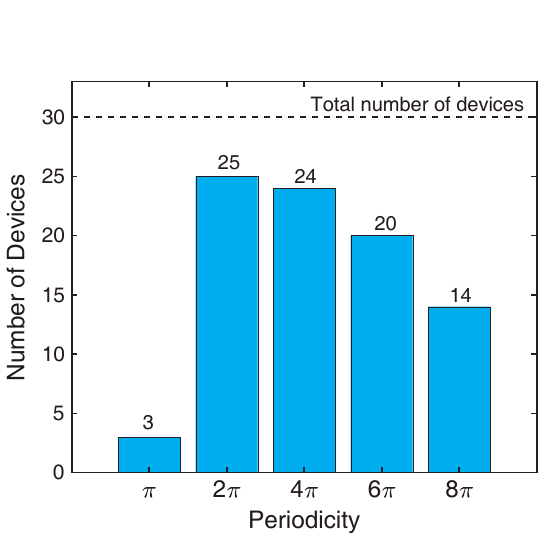}
	\caption{\label{fig5}
		\textbf{Histogram of the numbers of ring devices manifesting each periodicities.}
		The graph summarizes the experimental results obtained from a total number of 30 ring devices made of 70~nm-thick $\beta$-Bi$_2$Pd thin films epitaxially grown on SrTiO$_3$(001) and MgO(001) substrates with various device design geometries \cite{SM}.
	}
\end{figure}

\widetext
\clearpage
\begin{center}
	\textbf{\large Supplementary information for 
		fractional Little-Parks effect observed in a topological superconductor}
\end{center}

\setcounter{equation}{0}
\setcounter{figure}{0}
\setcounter{table}{0}
\setcounter{page}{1}
\makeatletter

\renewcommand{\thefigure}{S\arabic{figure}}
\renewcommand{\thetable}{S\arabic{table}}

\newcommand{\bs}[1]{{\boldsymbol{#1}}}
\newcommand{\red}[1]{{\textcolor{red}{#1}}}
\newcommand{\blue}[1]{{\textcolor{blue}{#1}}}
\newcommand{\magenta}[1]{{\textcolor{magenta}{#1}}}
\newcommand{\green}[1]{{\textcolor[rgb]{0,0.5,0}{#1}}}
\newcommand{\Tr}{\mathop{\mathrm{Tr}}}
\newcommand{\tr}{\mathop{\mathrm{tr}}}
\newcommand{\bsigma}{\boldsymbol{\sigma}}
\newcommand{\re}{\mathop{\mathrm{Re}}}
\newcommand{\im}{\mathop{\mathrm{Im}}}
\renewcommand{\b}[1]{{\mathbf{#1}}}
\newcommand{\diag}{\mathrm{diag}}
\newcommand{\sign}{\mathrm{sign}}
\newcommand{\sgn}{\mathop{\mathrm{sgn}}}
\renewcommand{\c}[1]{\mathcal{#1}}
\renewcommand{\d}{\text{\dj}}
\newcommand{\Res}{\mathop{\mathrm{Res}}}

\newcommand{\mb}{\bm}
\newcommand{\ua}{\uparrow}
\newcommand{\da}{\downarrow}
\newcommand{\ra}{\rightarrow}
\newcommand{\la}{\leftarrow}
\newcommand{\mc}{\mathcal}
\newcommand{\lra}{\leftrightarrow}
\newcommand{\nn}{\nonumber}
\newcommand{\half}{{\textstyle{\frac{1}{2}}}}
\newcommand{\mf}{\mathfrak}
\newcommand{\MF}{\text{MF}}
\newcommand{\IR}{\text{IR}}
\newcommand{\UV}{\text{UV}}
\newcommand{\Pf}{\mathop{\mathrm{Pf}}}

\textbf{Materials and Methods}

\begin{figure}
	\centering
	\includegraphics[width=14cm]{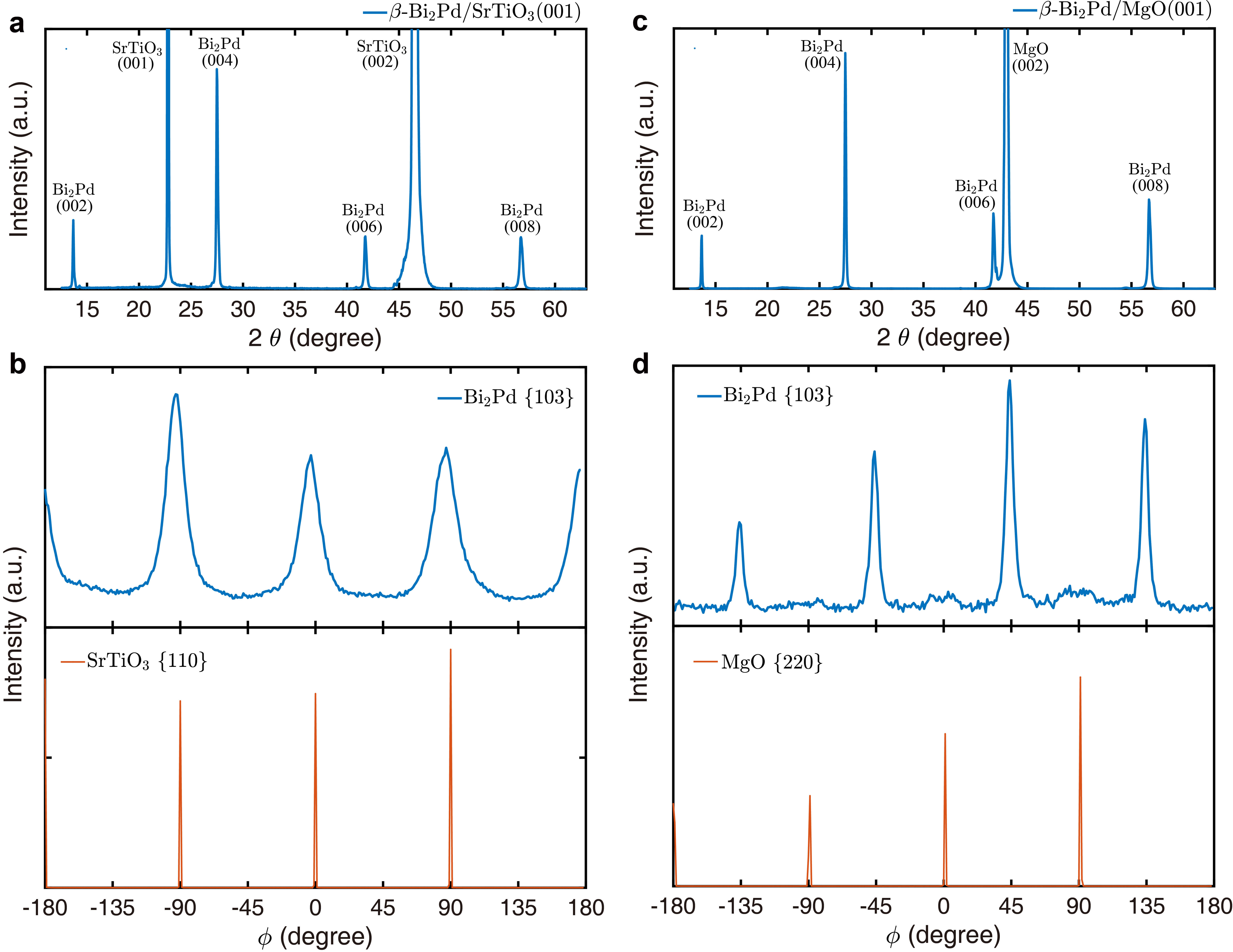}	
	\caption{
		\textbf{X-ray diffraction characterizations of epitaxial $\beta$-Bi$_2$Pd thin films.}
		\textbf{a}, $\theta$-$2\theta$ scan of $\beta$-Bi$_2$Pd/SrTiO$_3$(001). 
		\textbf{b}, In-plane $\phi$ scan of the $\beta$-Bi$_2$Pd $\{$103$\}$ planes (upper panel) and the SrTiO$_3$ $\{$110$\}$ planes (lower panel). 
		\textbf{c}, $\theta$-$2\theta$ scan of $\beta$-Bi$_2$Pd/MgO(001). 
		\textbf{d}, In-plane $\phi$ scan of the $\beta$-Bi$_2$Pd $\{$103$\}$ planes (upper panel) and the MgO $\{$220$\}$ planes (lower panel). 		
	}\label{fig:XRD}
\end{figure}

$\beta$-Bi$_2$Pd thin films were deposited by magnetron sputtering, with a base pressure of 3$\times$10$^{-8}$ Torr, onto substrates held at elevated temperature of 400 \textdegree C. 
The deposited films were capped by 1~nm-thick MgO protective layer, before taken out of the vacuum chamber. 
In this work, we prepared two different types of epitaxial $\beta$-Bi$_2$Pd thin films: $\beta$-Bi$_2$Pd/SrTiO$_3$(001) and $\beta$-Bi$_2$Pd/MgO(001). 
For both films, the $c$-axis of $\beta$-Bi$_2$Pd orients along the film normal direction. 
X-ray diffraction (XRD) results shown in Figs.~\ref{fig:XRD}a and ~\ref{fig:XRD}c reveal exclusively the (00$l$) peaks of $\beta$-Bi$_2$Pd. 
The epitaxial relations with respect to the two substrates, however, are different. 
Epitaxial $\beta$-Bi$_2$Pd/SrTiO$_3$, also reported previously by Lv \textit{et al.} \cite{Lv2017}, has $\beta$-Bi$_2$Pd[100]$\parallel$SrTiO$_3$[100]. 
On the other hand, $\beta$-Bi$_2$Pd rotates 45~\textdegree on MgO(001) and matches its [100] direction to MgO [110]. 
The $\phi$-scan demonstrated in Figs.~\ref{fig:XRD}b and \ref{fig:XRD}d confirm the epitaxial relations of film growth on SrTiO$_3$(001) and MgO(001), respectively.

\begin{figure}
	\centering
	\includegraphics[width=6cm]{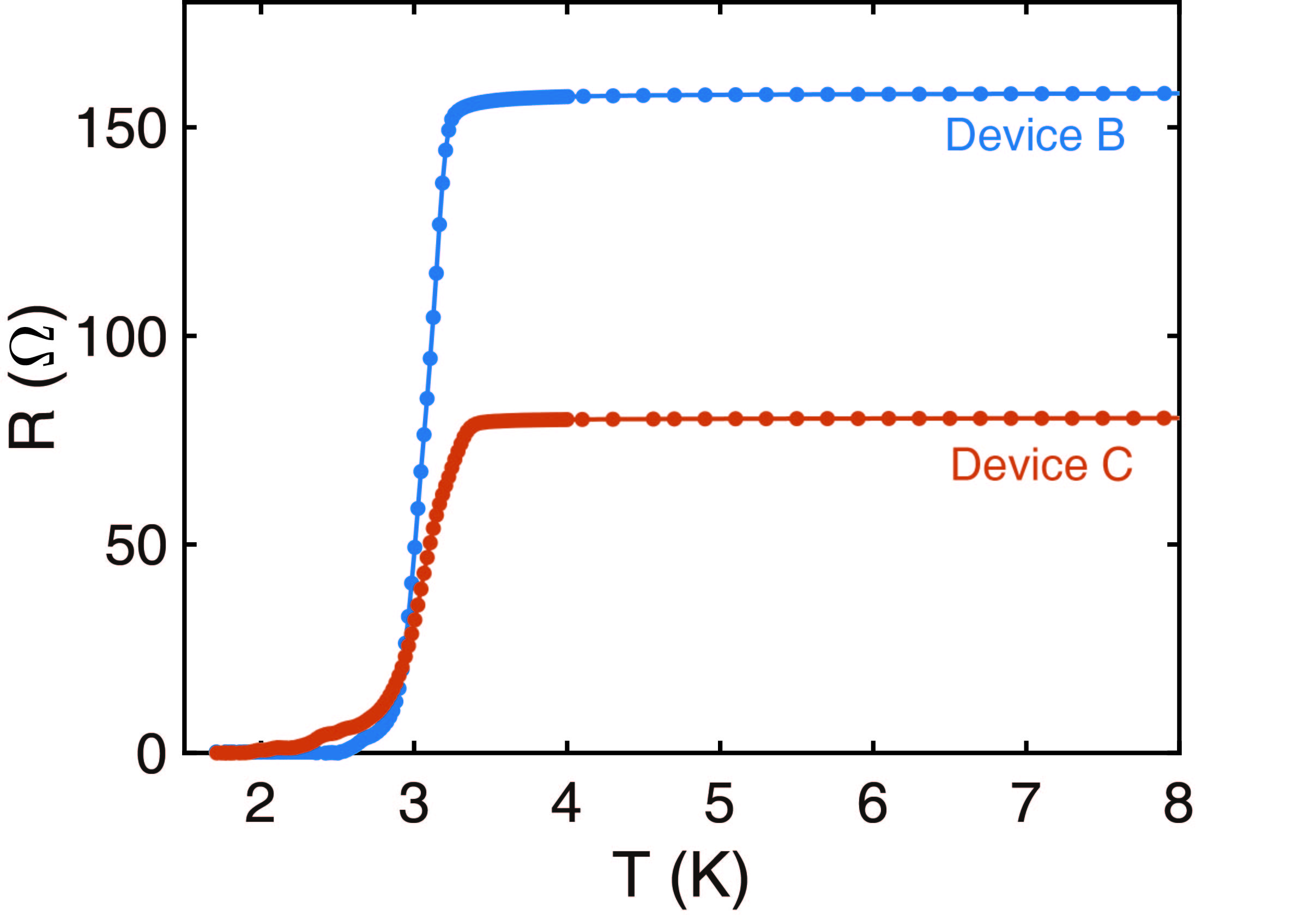}	
	\caption{
		\textbf{Temperature dependence of the electrical resistance}
		Resistance versus temperature curves for Device B (blue curve) and Device C (red curve). Device B is a type-I ring device of $\beta$-Bi$_2$Pd/MgO. Device C is a type-VI ring device of $\beta$-Bi$_2$Pd/SrTiO$_3$. 
	}\label{fig:RT}				
\end{figure}

The as-grown films both show transition temperatures of about 3.5~K, with a narrow transition of less than 0.2~K. 
The transition broadens in nano-ring devices \cite{Li2019}. 
Fig.~\ref{fig:RT} shows the R-T curves for Devices B, fabricated of $\beta$-Bi$_2$Pd/MgO(001) and the narrowest line width of 50~nm; and Device C, fabricated of  $\beta$-Bi$_2$Pd/SrTiO$_3$(001) and the narrowest line width of 400~nm. 
The temperature dependences are similar except for the difference in the absolute values of the normal state resistance. 
The Little-Parks effect is typically observed in a temperature window between 2~K and 3~K, similar to the case of polycrystalline $\beta$-Bi$_2$Pd/Si rings \cite{Li2019}. 
Above 3.2~K, no oscillations can be observed in magetoresistance. 

\begin{figure}
	\centering
	\includegraphics[width=18cm]{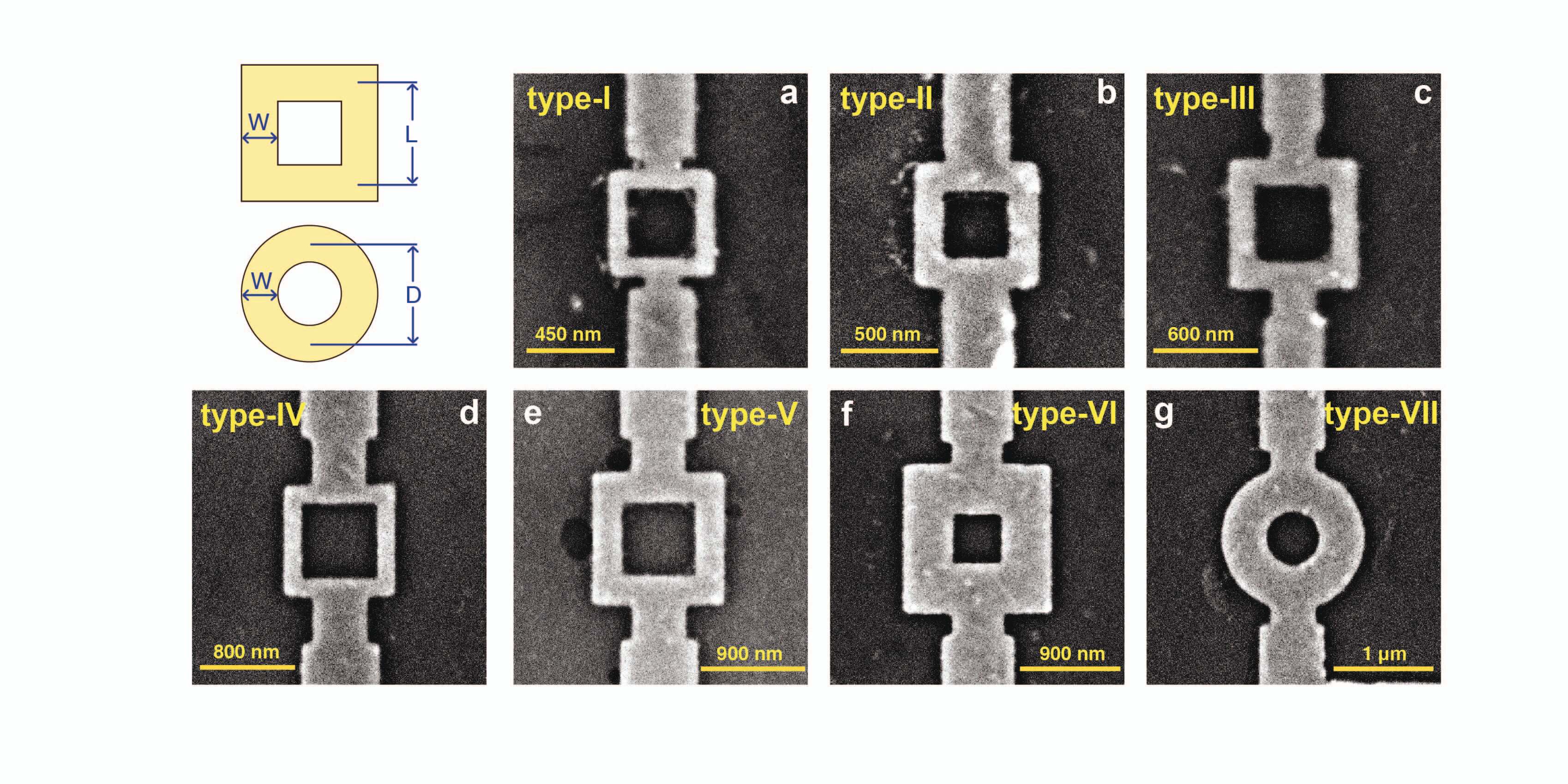}		
	\caption{
		\textbf{SEM images of $\beta$-Bi$_2$Pd ring devices.}
		For square-shaped rings, $L$ denotes the length between the midpoints of the opposing walls. 
		For round-shaped rings, $D$ denotes the diameter of the ring, also measured between the midpoints of the wall. 
		$W$ denotes the width of the wall. 
		We denote the geometric factors as [$L$, $W$] and \{$D$, $W$\}, for square-shaped and round-shaped designs, respectively. 
		The length of the scale bar in each figure corresponds to $L$ or $D$ of the ring device. 
		The sizes of various device designs are: 
		type-I [450 nm, 50 nm]; type-II [500 nm, 100 nm]; type-III [600 nm, 100 nm]; type-IV [800 nm, 100 nm]; type-V [900 nm, 200 nm]; type-VI [900 nm, 400 nm]; type-VII \{1~$\mu$m, 400 nm\}. 
	}\label{fig:Device}				
\end{figure}

The epitaxial thin films were fabricated into mesoscopic ring devices using a Vistec e-beam writer (EBPG5200ES). 
In this study we employed 7 different device designs. 
Fig.~\ref{fig:Device} demonstrates the scanning electron microscope (SEM) images of all 7 designs. 
They include square-shaped (type-I to VI) and round-shaped (type-VII) ring designs. 
The size of the square-shaped rings, $L$, measured as the distance between the midpoints of opposing walls, varies from 450~nm (type-I) to 900 nm (type-VI). 
The diameter ($D$) of the round-shaped type-VII design is 1~$\mu$m. 
The line width of the ring devices, $W$, varies from 50~nm (type-I) to 400~nm (type-VI and type-VII). 
In the rest of the text we denote the geometric factors as [$L$, $W$] for the square-shaped rings, and \{$D$, $W$\} for the round-shaped rings. 
Table~\ref{Tab:DeviceTable} summarizes the designs of the devices which are presented in the main text and later in the Supplementary Information.

\begin{table}[]
	\begin{tabular}{@{}|c|c|c|c|c|c@{}}
		\hline
		Device Index & Substrate & Design   & Expected $\Phi_0$-period (Oe) & Observed $\Phi_0$-period (Oe) &  \\
		\hline
		A            & STO       & type-IV  & 32.31                       & 33.50                       &  \\
		\hline
		B            & MgO       & type-I   & 102.12                      & 116.0                      &  \\
		\hline
		C            & STO       & type-VI  & 25.53                       & 27.56                       &  \\
		\hline
		D            & STO       & type-III & 57.44                       & 60.0                       &  \\
		\hline
		E			 & STO       & type-VI  & 25.53                       & 28.0                       &  \\
		\hline
		F			 & STO       & type-VII & 26.33                       & 26.0                        & \\
		\hline
		G			 & STO       & type-VII & 26.33                       & 22.50                        & \\
		\hline
		H			 & STO       & type-IV  & 32.31                       & 33.50                        & \\
		\hline
		I			 & STO       & type-III & 57.44                       & 52.80                        & \\
		\hline
		J			 & STO       & type-VI  & 25.53                       & 31.75                        & \\
		\hline
		K            & STO       & type-I   & 102.12                      & 121.0                      &  \\
		\hline
	\end{tabular}
	\caption{
		\textbf{Summary of all devices.}
	}
	\label{Tab:DeviceTable}
\end{table}

\textbf{Ordinary $2\pi$-periodic Little-Parks effect in polycrystalline $\beta$-Bi$_2$Pd and Nb devices}

\begin{figure}
	\centering
	\includegraphics[width=16cm]{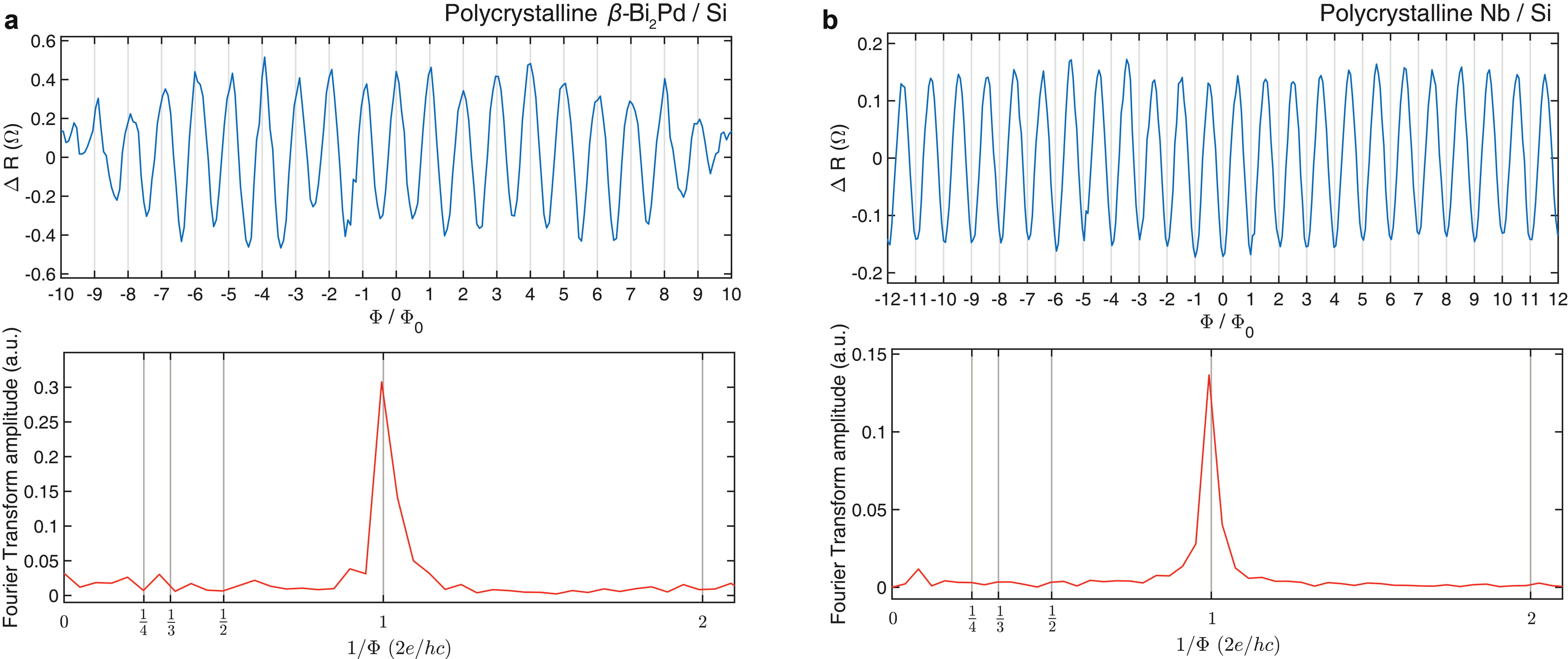}	
	\caption{
		\textbf{Flux quantization and $2\pi$-periodic Little-Parks effect.}
		\textbf{a}, Top panel: the Little-Parks oscillation observed in a ring device of polycrystalline $\beta$-Bi$_2$Pd~/~Si, reproduced from Ref.~\cite{Li2019}, geometric factors [900~nm, 200~nm]. Bottom panel: t                         he Fourier transform of the Little-Parks oscillation data in the top panel. 
		\textbf{b}, Top panel: the Little-Parks oscillation observed in a ring device of polycrystalline Nb~/~Si, reproduced from Ref.~\cite{Li2019}, geometric factors [800~nm, 100~nm]. Bottom panel: the Fourier transform of the Little-Parks oscillation data in the top panel. 
	}\label{fig:S2pi}				
\end{figure}

It is well known that flux quantization in a superconducting ring produces Little-Parks oscillations in period of $\Phi_0$. 
As examples, in Fig.~\ref{fig:S2pi} we reproduce the Little-Parks effect observed in polycrystalline $\beta$-Bi$_2$Pd and Nb ring devices, previously reported in Ref.~\cite{Li2019}. 
In the bottom panels, the Little-Parks oscillations are converted to the frequency domain by Fourier transform. 
The spectra clearly demonstrate the monochromatic $2\pi$-periodicity, in contrast to the fractional Little-Parks effect observed in epitaxial $\beta$-Bi$_2$Pd devices.

\textbf{Temperature dependence of the Little-Parks oscillations in Device A}

\begin{figure}
	\centering
	\includegraphics[width=16cm]{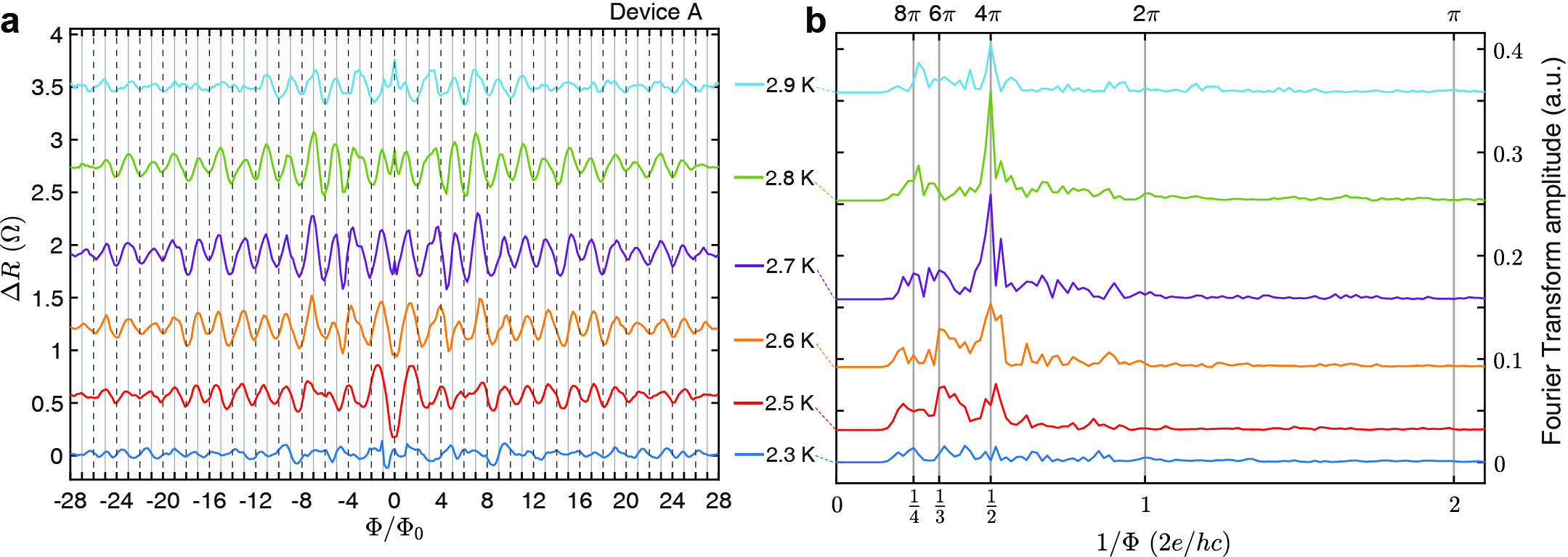}	
	\caption{
		\textbf{Temperature dependence of the Little-Parks effect of Device A. }
		\textbf{a}, The Little-Parks oscillation of Device A, a ring device of type-IV design [800~nm, 100~nm].
		\textbf{b}, The Fourier transform of the Little-Parks oscillation data in (\textbf{a}). Both figures share the same color code for various temperatures as shown in the middle.
	}\label{fig:SA_vT}				
\end{figure}

In the main text of the manuscript we presented the Little-Parks oscillation of Device A at a particular temperature of 2.7 K (Fig.~1). 
Here we present the results of the temperature dependence from 2.3~K to 2.9~K in Fig.~\ref{fig:SA_vT}, where the Little-Park effect can be observed. 
Generally, the $4\pi$-component is the dominating periodicity whereas a sizable $6\pi$-component can be found at lower temperatures. 

\textbf{Analyzing the oscillation periods by fitting}

\begin{figure}
	\centering
	\includegraphics[width=9cm]{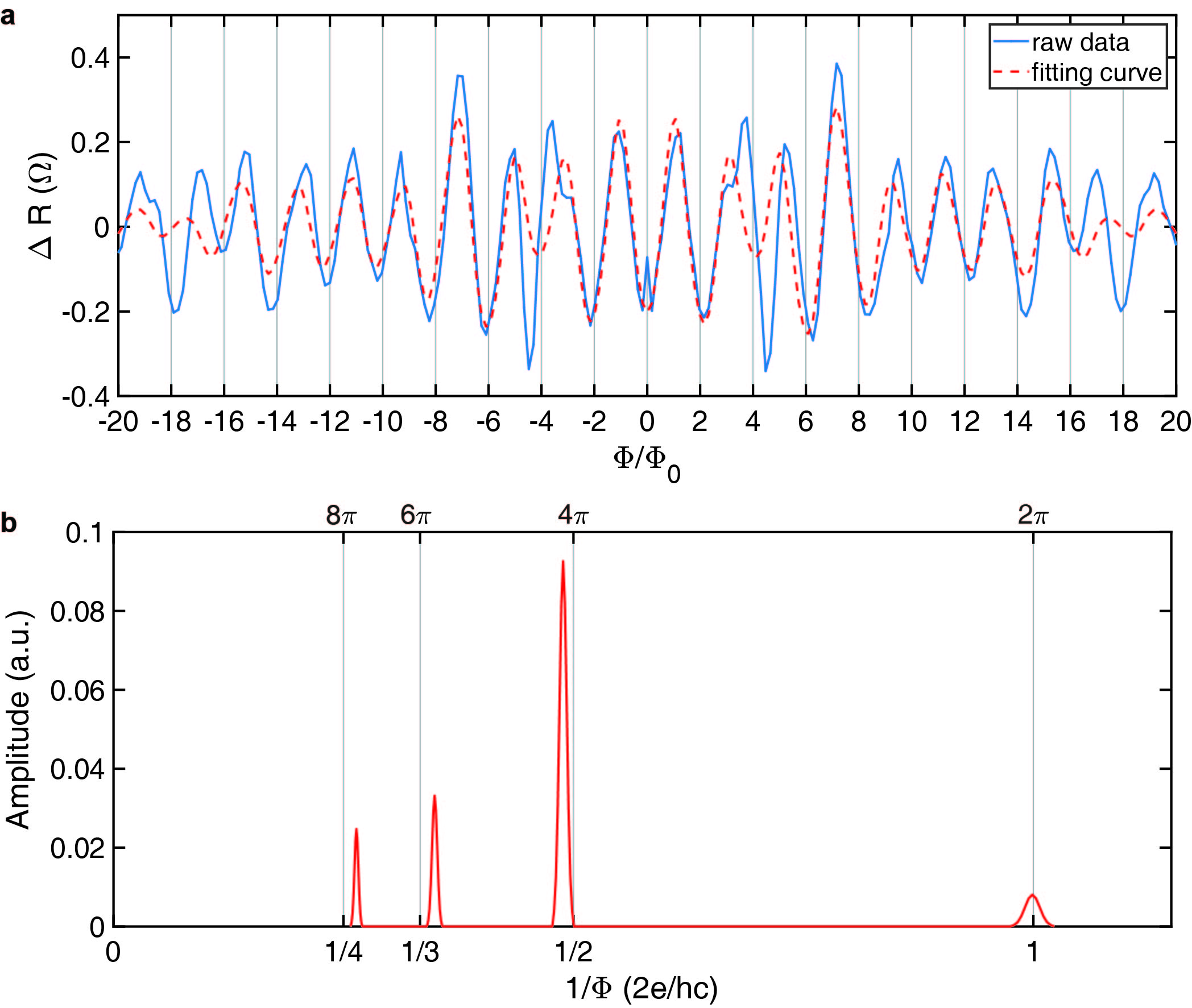}	
	\caption{
		\textbf{Fitting the Little-Parks oscillation of Device A. }
		\textbf{a}, The fitting result to the data presented in Figure~1d. The fitting allows four cosine terms corresponding to $2\pi$-, $4\pi$-, $6\pi$-, $8\pi$-periodicities, respectively. 
		\textbf{b}, The relative amplitude of the four periodicities derived by the fitting shown in (\textbf{a}). 
	}\label{fig:Sfitting}				
\end{figure}

Other than Fourier transform, an alternative approach for analyzing the oscillation periods is by fitting the magnetoresistance raw data employing a series of cosine functions with discrete frequencies. 
We analyze the Little-Parks effect of Device A, the data presented in Fig.~1d. 
We fit the raw data with a function which is the sum of four cosine terms with discrete frequencies, corresponding to $2\pi$-, $4\pi$-, $6\pi$- and $8\pi$-periodicities, respectively. 
The amplitude and the frequency of each term are the fitting parameters. 
The result is presented in Fig.~\ref{fig:Sfitting}. 
The fitting provides a cleaner spectrum (Fig.~\ref{fig:Sfitting}b), which is nevertheless consistent with the Fourier transform result presented in Fig.~1e. 
The $4\pi$-periodicity dominates the Little-Parks oscillation. 
Furthermore, the fitting result also reveals the ordinary $2\pi$-component whose amplitude is too small to be distinguished from the noise floor in the Fourier transform spectrum shown in Fig.~1e. 

\textbf{Analysis of the FFT spectrum}

To study the oscillation periods of the Little-Parks effect, we performed fast Fourier transform (FFT) analysis on the magnetoresistance raw data. 
The results have been shown in the main text as well as here in the Supplementary Materials, which could reveal multiple coexisting periodicities. 

However, if the oscillatory amplitude of a specific frequency is small, it may become difficult to distinguish the signal from the noise floor of the FFT spectrum, even though features of the said frequency may be visible from the $\Delta R$-$\Phi$ raw data. 
One such example is the $2\pi$-periodicity of Device A. 
While the presence of the $\Phi_0$-period produces visible modulations to the Little-Parks oscillation such as the kink-like feature at $\Phi=\pm$3$\Phi_0$, it is difficult to find corresponding 2$\pi$-peaks in the FFT spectra (Fig.~1e and Fig.~\ref{fig:SA_vT}b). 

\begin{figure}
	\centering
	\includegraphics[width=16cm]{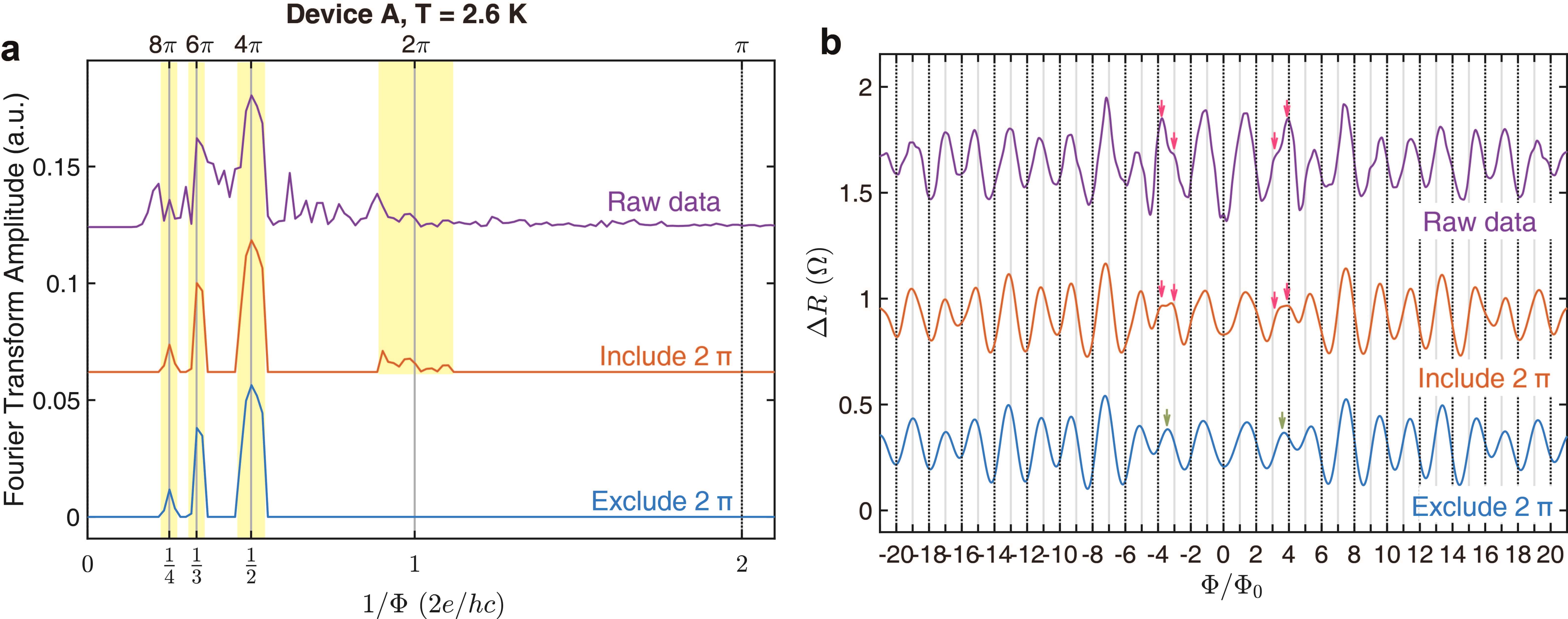}	
	\caption{
		\textbf{FFT analysis of the $2\pi$-periodicity in Device A.}
		\textbf{a}, The Fourier transform spectra employed in the analysis. The purple curve is the spectrum converted from the experimental data. The red curve is the filtered spectrum allowing the spectral weight near the $2\pi$-, $4\pi$-, $6\pi$- and $8\pi$-periodicities; while the blue curve excludes the $2\pi$-periodicity. The yellow shade denotes the filter window. 
		\textbf{b}, Little-Parks oscillation that is experimentally observed (purple) or reconstructed from the filtered FFT spectrum (red and blue). The reconstructed oscillations either include (red) or exclude (blue) the $2\pi$-periodic component. 
	}\label{fig:SA_FFT}				
\end{figure}

In such cases, we further the FFT analysis in order to separate the signal of these periodicities from the noise.  
First, we create filters to only allow the spectral weight around a number of discrete integer-quantum frequencies ($2\pi$, $4\pi$, \textit{etc.}), blocking the spectral continuum of the noise. 
As an example, we present the analysis for the Little-Parks oscillation of Device A at 2.6~K in Fig.~\ref{fig:SA_FFT}a. 
Each filter window for a specific frequency, marked by the yellow-shaded area in Fig.~\ref{fig:SA_FFT}a, can be independently turned on and off. 
By excluding and including the $2\pi$-periodicity, we obtain the blue and the red spectral curves, respectively. 
We shall then reconstruct the Little-Parks oscillations ($\Delta R$ vs. $H$) from the filtered spectra, as shown in Fig.~\ref{fig:SA_FFT}b, and compare them with the experimental raw data (purple curve). 
Apparently, only by including the spectral weight around the $2\pi$-periodicity can one reproduce the oscillatory features in the raw data marked by the pink arrows. 
The analysis confirms that the kink-like features at $\pm$3$\Phi_0$ are indeed the result of the $2\pi$-periodic oscillation. 

\begin{figure}
	\centering
	\includegraphics[width=16cm]{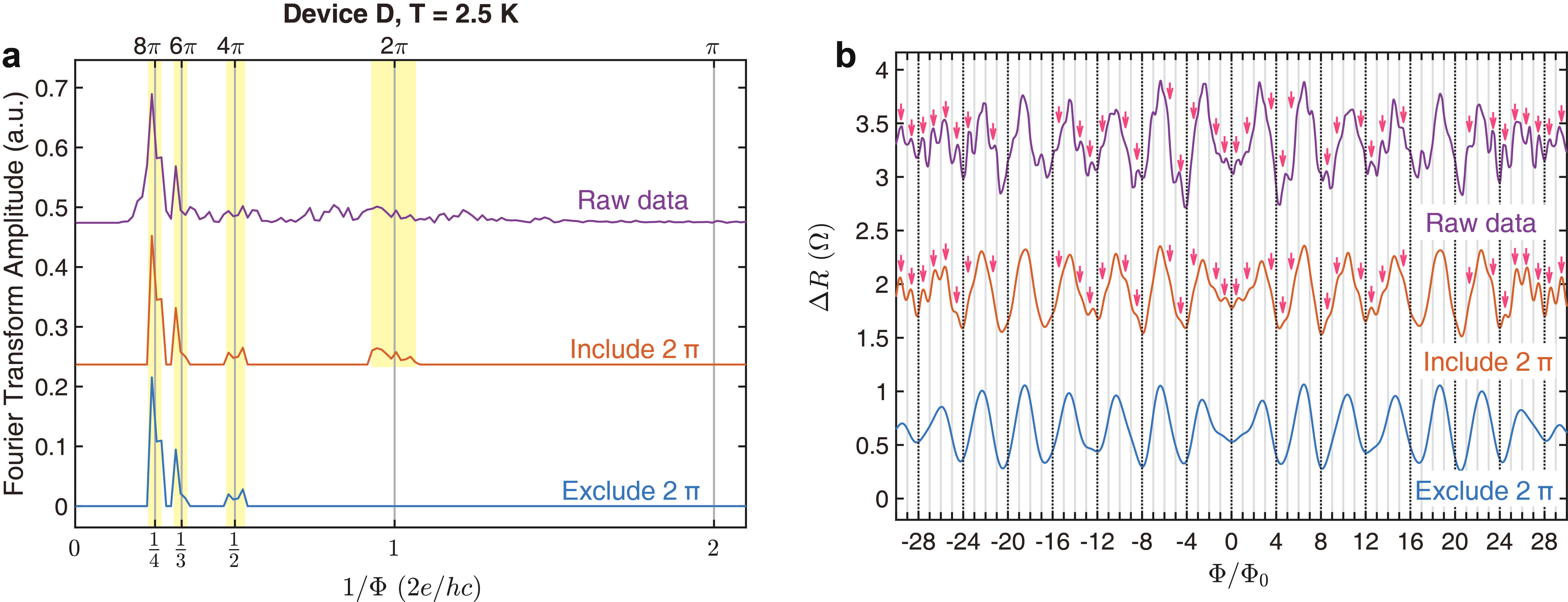}	
	\caption{
		\textbf{FFT analysis of the $2\pi$-periodicity in Device D.}
		\textbf{a}, The Fourier transform spectra employed in the analysis. The purple curve is the spectrum converted from the experimental data. The red curve is the filtered spectrum allowing the spectral weight near the $2\pi$-, $4\pi$-, $6\pi$- and $8\pi$-periodicities; while the blue curve excludes the $2\pi$-periodicity. The yellow shade denotes the filter window. 
		\textbf{b}, Little-Parks oscillation that is experimentally observed (purple) or reconstructed from the filtered FFT spectrum (red and blue). The reconstructed oscillations either include (red) or exclude (blue) the $2\pi$-periodic component. 		
	}\label{fig:SD_FFT}				
\end{figure}

In Fig.~\ref{fig:SD_FFT}, we present another example of the FFT analysis, concerning the presence of the $2\pi$-periodicity in Device D. 
Apparently the $2\pi$-periodic component is present in the $\Delta R$-$\Phi$ raw data as the high-pitch oscillations shown in Fig.~4a of the main text. 
However its relatively small amplitude makes the corresponding $2\pi$-peaks rather invisible in the FFT spectra. 
Here we employ the same FFT analysis by creating filtered spectra with or without the $2\pi$ spectral weight (Fig.~\ref{fig:SD_FFT}a). 
Reconstructed from the filtered spectrum excluding the $2\pi$ component, the blue curve shown in Fig.~\ref{fig:SD_FFT}b fails to reproduce the high-pitch oscillatory features in the raw data, marked by pink arrow on the purple curve. 
These features, on the other hand, are reproduced in the reconstructed Little-Parks oscillation by adding the filter window at $2\pi$ (red curve). 
Therefore we confirm the presence of the $2\pi$-periodicity in Device D. 

The FFT analyses also show that the blocked continuous noise spectra are unimportant, as the filtered spectra containing only the quantized frequencies reproduce the experimental observation reasonably well. 
In both cases the reconstructed oscillations (red curves) retain the essential features of the original experimental data. 
The remaining deviations are mostly in the quantitative details, accounting for the blocked spectral weight of the continuum domains. 
Such deviation may be introduced by the background subtraction employed to obtain $\Delta R$, which is inevitably insufficient because the precise analytical expression of the background curve is difficult to configure. 

\textbf{Half-quantum $\pi$-periodicity}

\begin{figure}
	\centering
	\includegraphics[width=16cm]{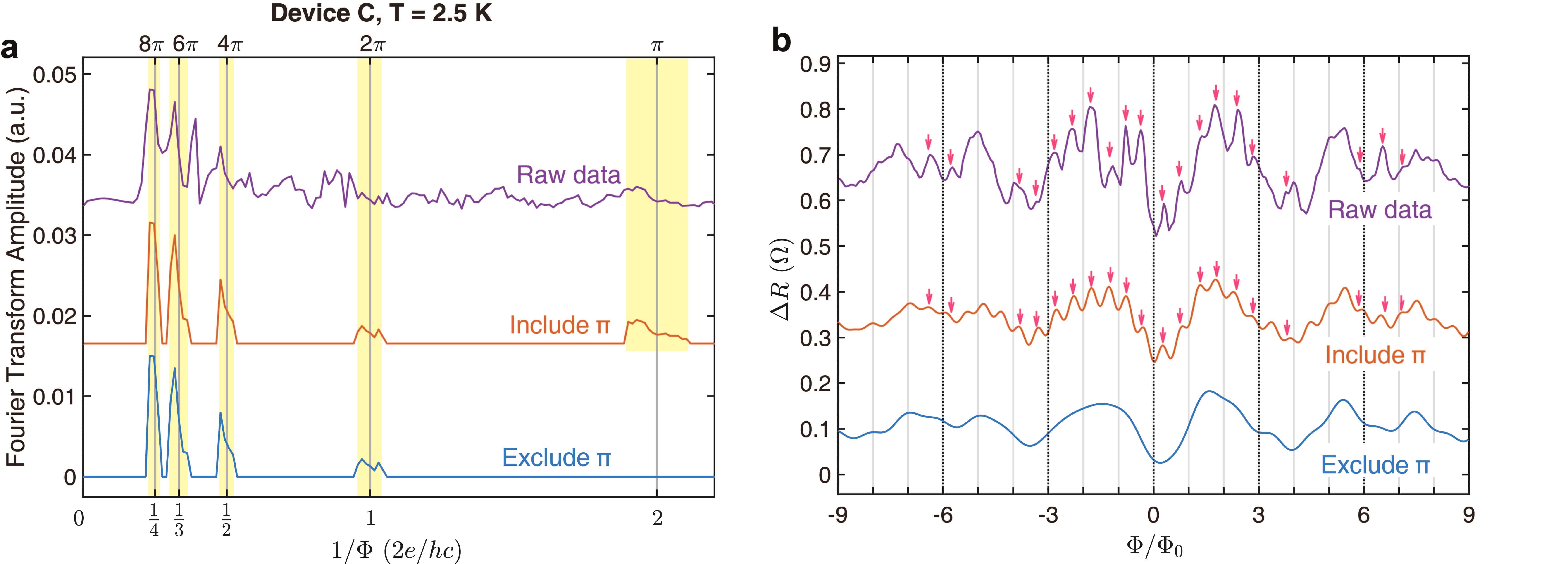}	
	\caption{
		\textbf{FFT analysis of the $\pi$-periodicity in Device C.}
		\textbf{a}, The Fourier transform spectra employed in the analysis. The purple curve is the spectrum converted from the experimental data. The red curve is the filtered spectrum allowing the spectral weight near the $\pi$-, $2\pi$-, $4\pi$-, $6\pi$- and $8\pi$-periodicities; while the blue curve excludes the $\pi$-periodicity. The yellow shade denotes the filter window. 
		\textbf{b}, Little-Parks oscillation that is experimentally observed (purple) or reconstructed from the filtered FFT spectrum (red and blue). The reconstructed oscillations either include (red) or exclude (blue) the $\pi$-periodic component. 
	}\label{fig:SC_FFT}				
\end{figure}

In this section we focus on the Little-Parks oscillation of Device C at 2.5~K which, as we have briefly discussed in the main text, manifests a $\pi$-periodic oscillation. 
These high-pitch oscillations with the period of $\frac{1}{2}\Phi_0$ are clearly present as can be seen from the raw data (purple curve in Fig.~\ref{fig:SC_FFT}b), however their small magnitude prohibits distinguishing the corresponding peak from the FFT spectrum (Fig.~\ref{fig:SC_FFT}a). 
We perform the FFT analysis as introduced in the previous section. 
The $\pi$-periodic oscillation features, marked by pink arrows in Fig.~\ref{fig:SC_FFT}b, can be reproduced from the filtered spectrum allowing the spectral weight around the half-quantum frequency of $\frac{2}{\Phi_0}$. 

It has long been anticipated that the spin-triplet superconductor or superfluid may host half-quantum vortices (HQV) in the equal-spin-pairing state \cite{volovik_line_1976,Mackenzie2003}. 
Experimental signatures of HQV have been reported for $^3$He  \cite{autti_observation_2016} and Sr$_2$RuO$_4$ \cite{jang_observation_2011,yasui_little-parks_2017}. 
For the case of Sr$_2$RuO$_4$, an in-plane magnetic field in addition to the perpendicular field is required; otherwise only integer-quantum quantization can be observed. 
The magnitude of the in-plane field usually is much greater than that of the perpendicular field \cite{jang_observation_2011,yasui_little-parks_2017}. 
In our case, the half-quantum oscillation in $\beta$-Bi$_2$Pd is spontaneous upon applying the perpendicular field, with no in-plane field. 

For $\beta$-Bi$_2$Pd, topological and spin-chiral surface state has been revealed by photo-emission \cite{Sakano2015} and quasi-particle interference imaging \cite{Iwaya2017}. 
Together with anisotropic superconducting pairing symmetry evidence by flux quantization \cite{Li2019}, the experimental results strongly suggest $p$-wave spin-triplet pairing. 
$\beta$-Bi$_2$Pd has a tetragonal crystalline structure and a cylindrical Fermi surface \cite{Sakano2015}, much resemble that of Sr$_2$RuO$_4$. 
Future works are required to fully resolve the pairing symmetry of $\beta$-Bi$_2$Pd. 
It would be particularly intereting to examine if the many $p$-wave states considered for Sr$_2$RuO$_4$ \cite{Mackenzie2003,Kallin2012} may also be possible for $\beta$-Bi$_2$Pd, especially the chiral $p$-wave state which is proposed to give rise to the HQV.

\textbf{More examples of non-$2\pi$-periodicities}

\begin{figure}
	\centering
	\includegraphics[width=16cm]{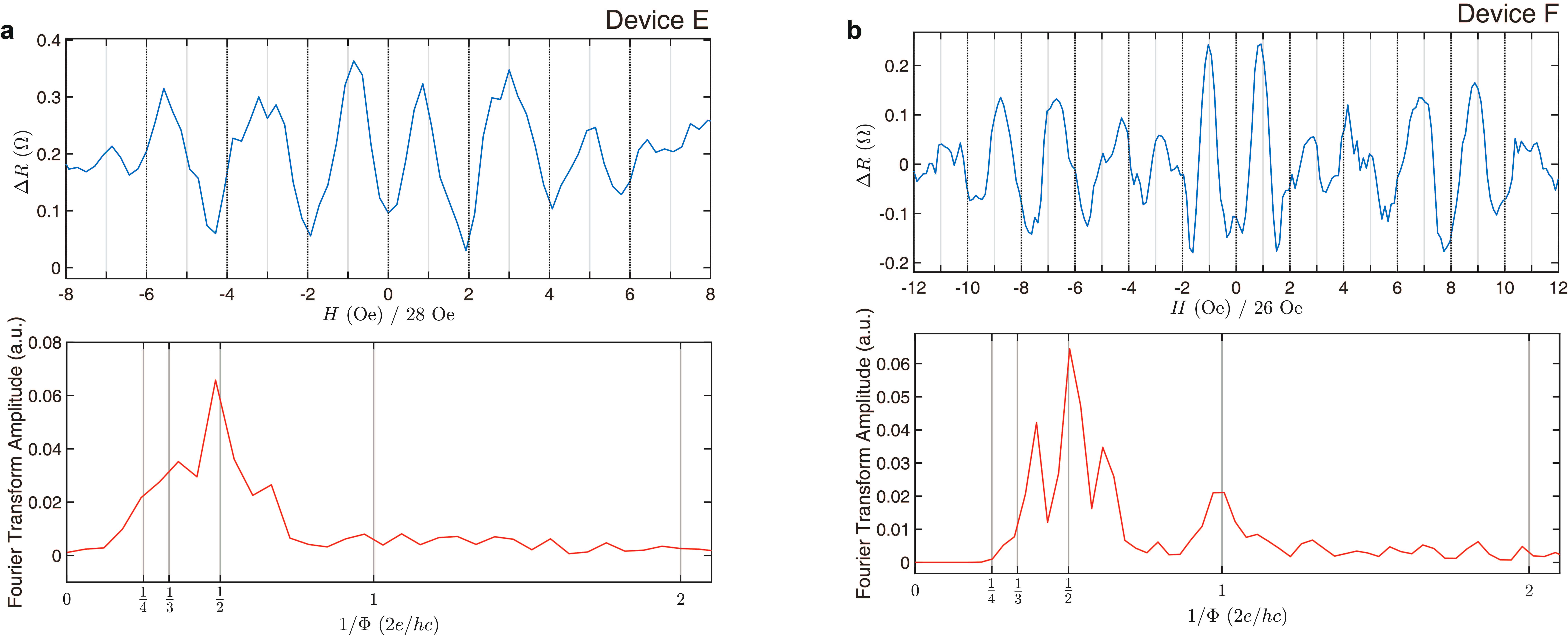}	
	\caption{
		\textbf{Examples of $4\pi$-periodicity.}
		\textbf{a}, The Little-Parks effect (upper panel) and the FFT spectrum (lower panel) of Device E at 2.5~K. The device design of Device E is type-VI [900~nm, 400~nm]. 
		\textbf{b}, The Little-Parks effect (upper panel) and the FFT spectrum (lower panel) of Device F at 2.3~K. The device design of Device F is type-VII \{1~$\mu$m, 400 nm\}. 		
	}\label{fig:SC_4pi}				
\end{figure}

\begin{figure}
	\centering
	\includegraphics[width=16cm]{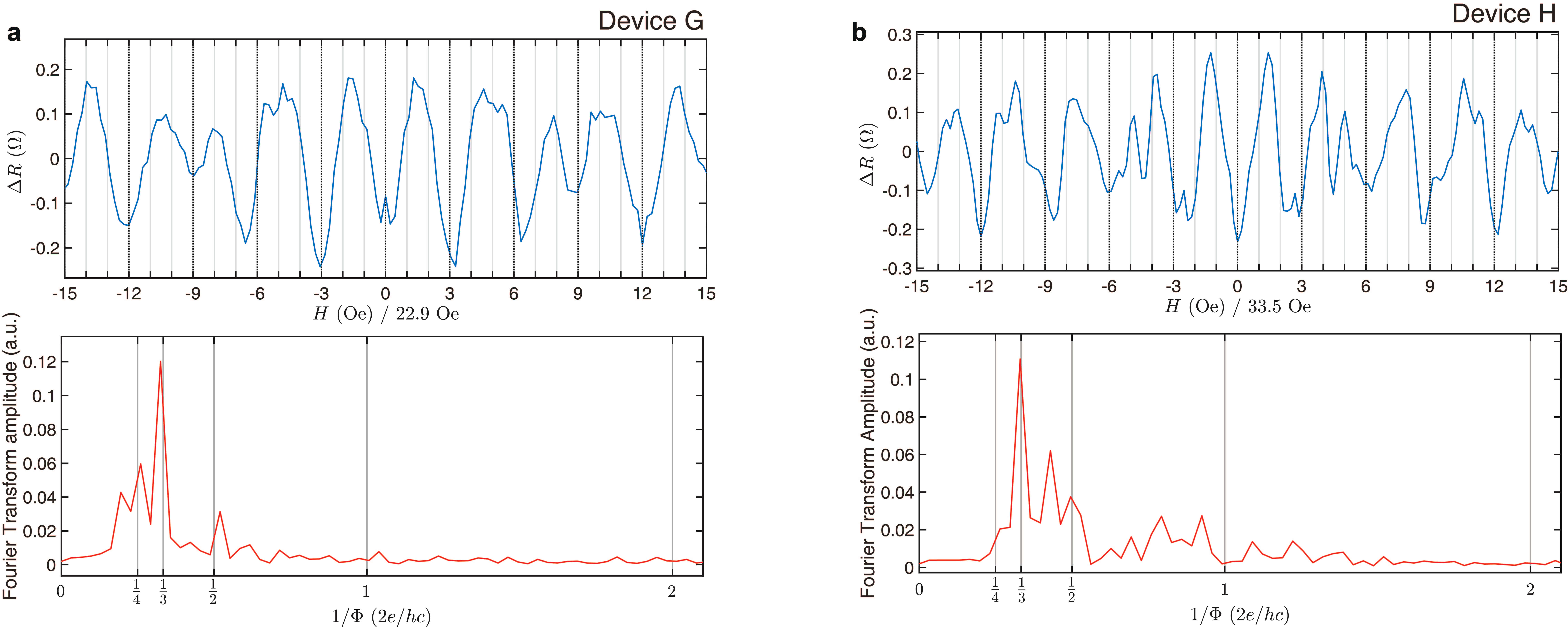}	
	\caption{
		\textbf{Examples of $6\pi$-periodicity.}
		\textbf{a}, The Little-Parks effect (upper panel) and the FFT spectrum (lower panel) of Device G at 2.75~K. The device design of Device G is type-VII \{1~$\mu$m, 400 nm\}. 
		\textbf{b}, The Little-Parks effect (upper panel) and the FFT spectrum (lower panel) of Device H at 2.5~K. The device design of Device H is type-IV [800~nm, 100~nm]. 
	}\label{fig:SC_6pi}				
\end{figure}

\begin{figure}
	\centering
	\includegraphics[width=16cm]{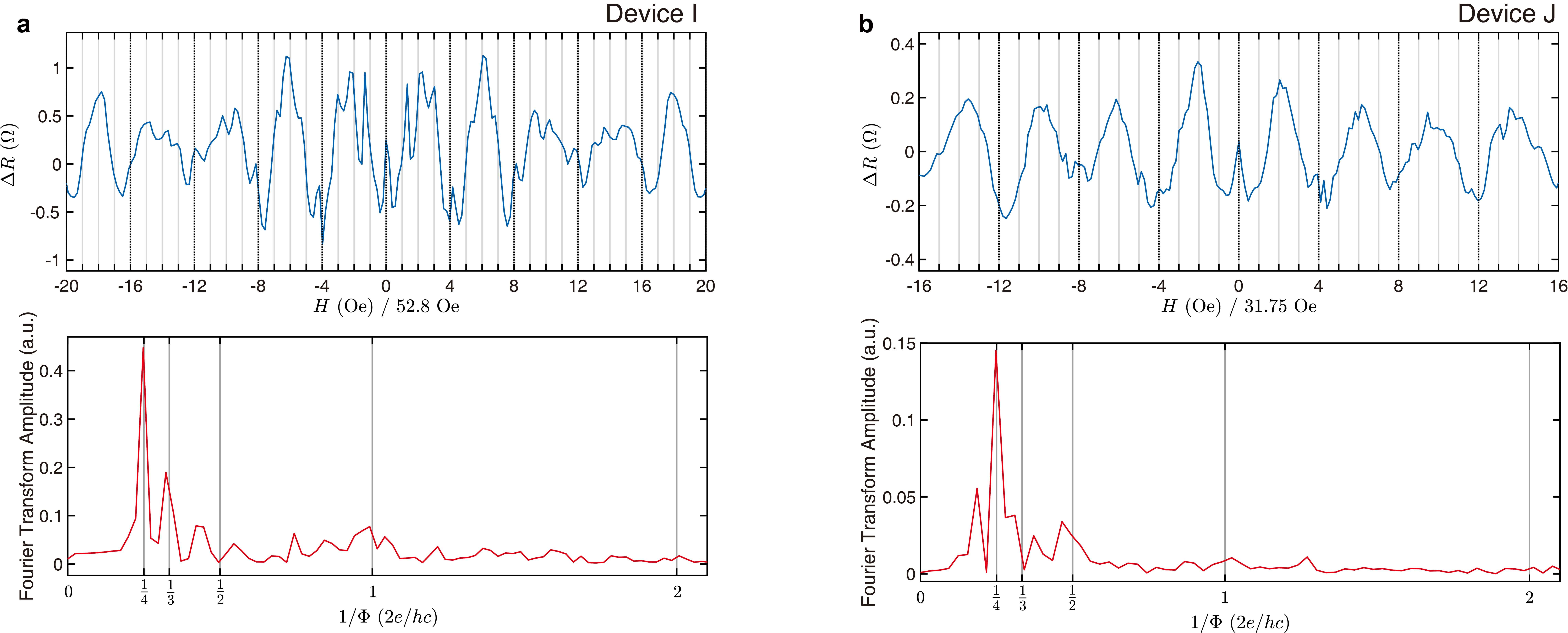}	
	\caption{
		\textbf{Examples of $8\pi$-periodicity.}
		\textbf{a}, The Little-Parks effect (upper panel) and the FFT spectrum (lower panel) of Device I at 2.4~K. The device design of Device I is type-III [600~nm, 100~nm]. 
		\textbf{b}, The Little-Parks effect (upper panel) and the FFT spectrum (lower panel) of Device J at 2.7~K. The device design of Device J is type-VI [900~nm, 400~nm]. 
	}\label{fig:SC_8pi}				
\end{figure}

In this section we present additional examples of devices manifesting fractional Little-Parks effect. 
For each non-$2\pi$-periodicities ($4\pi$, $6\pi$ and $8\pi$), we show two more devices whose Little-Parks oscillations are dominated by the said periodicity (Figs.~\ref{fig:SC_4pi}, \ref{fig:SC_6pi} and \ref{fig:SC_8pi}). 
These devices are labeled with letters E to J. 
Information about the device design and the $\phi_0$-oscillation period is also summarized in Table~\ref{Tab:DeviceTable}. 

It can be seen that the device design does not appear to be a deciding factor as to which anomalous periodicities shall be observed.
For example, Device H shares the same design with Device A. 
However the former shows primarily the $6\pi$-periodicity whereas the latter shows overwhelmingly the $4\pi$-periodicity. 

\textbf{Magnetoresistance anomaly at zero magnetic field.}

\begin{figure}
	\centering
	\includegraphics[width=7cm]{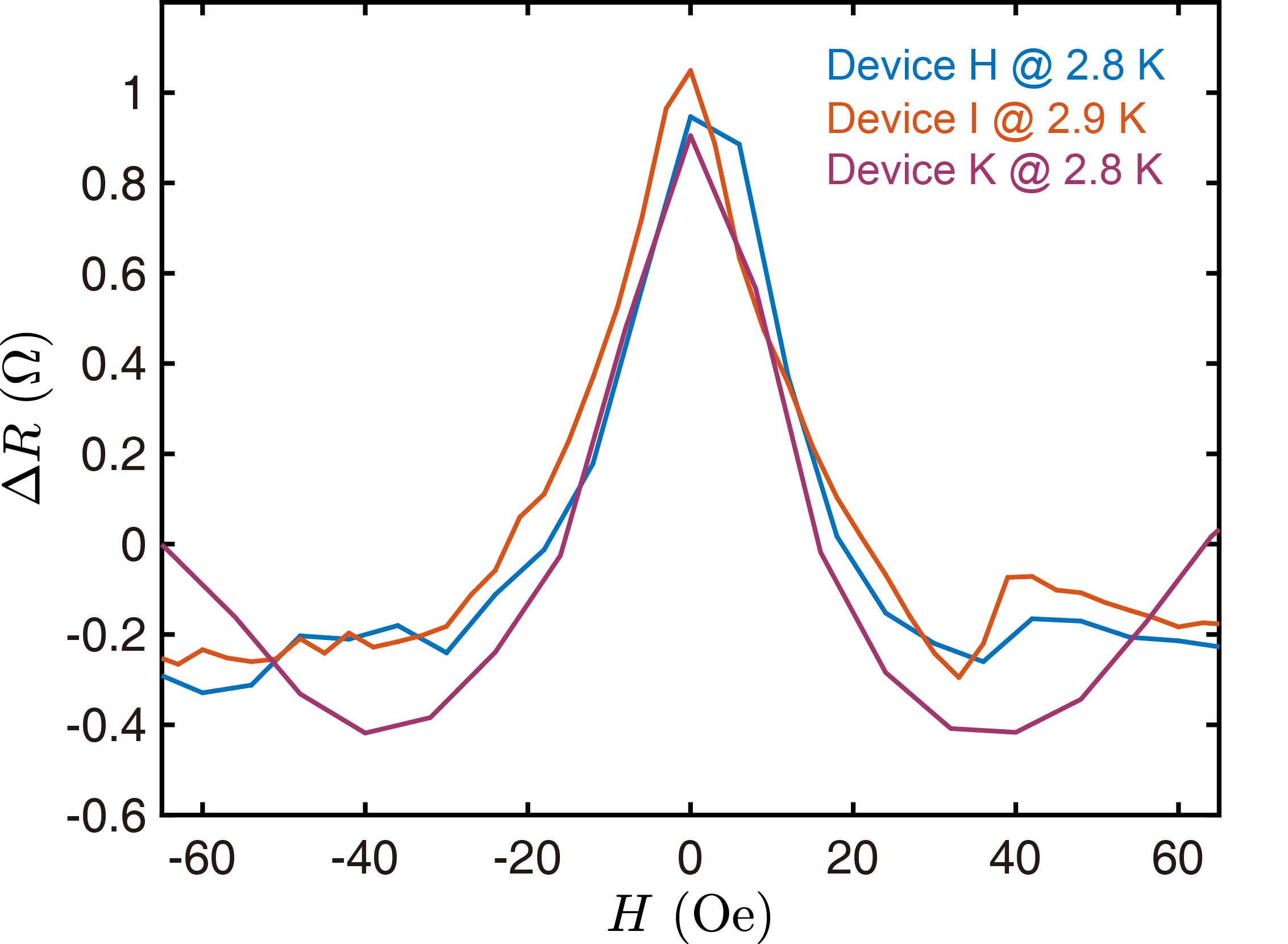}	
	\caption{
		\textbf{Zero-field magnetoresistance anomaly.}
		The magnetoresistance data of ring devices of three different designs. Device H (blue curve) is type-IV [800~nm, 100~nm] with $\Phi_0$-period of 33.5~Oe. Device I (red curve) is type-III [600~nm, 100~nm] with $\Phi_0$-period of 52.8~Oe. Device K (purple curve) is type-I [450~nm, 50~nm] with $\Phi_0$-period of 121~Oe. 
	}\label{fig:WL}				
\end{figure}

For a number of devices, the magnetoresistance demonstrates a spike-like anomaly around zero magnetic field, in addition to the Little-Parks oscillation. 
This behavior usually does not present itself at lowest temperatures, but becomes more prominent at higher temperatures ($>$2.7~K). 
Such behavior can be seen, for example, in the temperature dependence of the magnetoresistance in Devices A and D, presented in Fig.~\ref{fig:SA_vT}a and Fig.~4a. 
This suggests that the zero-field spike is related to the normal metallic state rather than the superconducting state. 
In fact, its rapidly decaying behavior at finite magnetic field resembles that of weak localization, known to be present in metals with sizable spin-orbit interaction \cite{Bergmann1984}. 
Observations of the localization effect in superconductors have also been reported \cite{Bruynseraede1983,Bergmann1984a,Hikita1990}. 
At any rate, this spike-like feature is not likely to be related to the magnetic flux quantization in the ring structure. 
Unlike the Little-Parks magnetoresistance oscillation, the field dependence of the zero-field spike is not determined by the dimensions of the rings. 
In Fig.~\ref{fig:WL} we present the most pronounced zero-field spike signals observed in three devices with various form factors. 
Because of the different ring sizes, the $\Phi_0$ oscillation periods varies by a factor of 4 (33.5~Oe of Device H versus 121~Oe of Device K). 
Yet the field dependence of the spike-like features are virtually identical for all three devices. 
Therefore the origin of the zero-field spike trace back to the properties of the bulk magnetoresistance.  
Although this phenomenon invites further investigations, we conclude that it does not concern the investigation of the Little-Parks effect.

\textbf{Fractional Little-Parks effect in a Kitaev chain.}

In this section we study theoretically the free energy of a Kitaev chain with a single Josephson junction oscillating in flux with the periodicity equal to two flux quanta. 
The $2\Phi_0$ periodicity comes from the single-electron tunneling via Majorana modes. 
The single-electron tunneling leads to the resistance of a $p$-wave superconductor at the critical temperature doubles its oscillation periodicity comparing to the conventional $\Phi_0$-periodic flux quantization. 

We start with the Hamiltonian of a Kitaev chain\cite{Kitaev_2001} to model a 1D $p$-wave superconducting ring with one single Josephson Junction \cite{AliceaBraiding} as

\begin{eqnarray}
H_{\alpha}=t\sum^{N-1}_{x=1} &&(e^{-i\phi_{\alpha}/2}c^{\dagger}_{\alpha,x}+e^{i\phi_{\alpha}/2}c_{\alpha,x})\times\nonumber\\
&&(e^{-i\phi_{\alpha}/2}c^{\dagger}_{\alpha,x+1}-e^{i\phi_{\alpha}/2}c_{\alpha,x+1})
\end{eqnarray}
Here $\alpha=L/R$ denotes the superconductors on left and right side of a Josephson Junction. $\phi_{\alpha}$ and $c$ are the superconducting phase and the electron operator respectively. The phase difference $\phi=\phi_L-\phi_R=2\pi(\Phi_B/\Phi_0)$ is defined by the flux $\Phi_B$ through the ring and flux quantum $\Phi_0=hc/(2e)$.

We model the single-electron tunneling across the Josephson junction as
\begin{eqnarray}
H_{\Gamma}=-\Gamma(c^{\dagger}_{L,N}c_{R,1}+h.c.)
\end{eqnarray}
By expressing the electron operators in terms of Majorana operators $c_{\alpha,x}=e^{-i\phi_{\alpha}/2}(\gamma^{\alpha}_{B,x}+i\gamma^{\alpha}_{A,x})/2$, and defining $d_{\alpha,x}=(\gamma^{\alpha}_{A,x+1}+i\gamma^{\alpha}_{B,x})/2$, we diagonalize the Kitaev chian $H_{\alpha}$ as
\begin{eqnarray}
H_{\alpha}=t\sum_{x=1}^{N-1}(2d^{\dagger}_{\alpha,x} d_{\alpha,x}-1)
\end{eqnarray}
The tunneling across the Josephson junction $H_{\Gamma}$ becomes
\begin{eqnarray}
&&H_{\Gamma}=\frac{\Gamma}{2}\{\cos(\phi/2)[d^{\dagger}_{R,1}(d^{\dagger}_{L,N-1}+d_{L,N-1})+h.c.]\nonumber\\
&&+\sin(\phi/2)[id^{\dagger}_{end}(d^{\dagger}_{R,1}-d_{R,1}+d^{\dagger}_{L,N-1}+d_{L,N-1})+h.c.]\nonumber\\
&&+\cos(\phi/2)[2d^{\dagger}_{end} d_{end}-1]\}.
\end{eqnarray}
Where $\phi=\phi_L-\phi_R$. The tunneling Hamiltonian can be simplified furthermore if we transform the basis into
\begin{eqnarray}
d_A&=&\frac{1}{\sqrt{2}}(d_{L,N-1}+d_{R,1})\\
d_B&=&\frac{1}{\sqrt{2}}(d_{L,N-1}-d_{R,1}).
\end{eqnarray}
The tunneling Hamiltonian in such basis becomes
\begin{eqnarray}
H_{\Gamma}=&&\frac{\Gamma}{2}\{\cos(\phi/2)[(2d^{\dagger}_{end} d_{end}-1)+(d^{\dagger}_A d_A-d^{\dagger}_B d_B)]\nonumber\\
&&+\cos(\phi/2)(d^{\dagger}_{A}d^{\dagger}_{B}+d_{B}d_{A})\nonumber\\
&&+\sqrt{2}\sin(\phi)[id^{\dagger}_{end}(d^{\dagger}_A+d_B)+h.c.]\}.
\end{eqnarray}
The relevant terms in Kitaev chain $H_{\alpha}$ is given as $H_{eff}$, where
\begin{eqnarray}
H_{eff}=t[(2d^{\dagger}_A d_A-1)+(2d^{\dagger}_B d_B-1)].
\end{eqnarray}

By defining $\psi=(d_{end},d_B,d_A^{\dagger})$, we can write the BdG Haniltonian $H_{BdG}=H_{eff}+H_{\Gamma}$ in the basis of $\Psi=[\psi,(\psi^{\dagger})^T]^T$ as $H_{BdG}=\Psi^{\dagger}\hat{H}_{BdG}\Psi$. Here $\hat{H}_{BdG}$ is a $6\times 6$ matrix and can be block diagonal into two $3\times 3$ matrices as
\begin{eqnarray}
\hat{H}_{BdG}=\begin{bmatrix}
\hat{h}_{BdG}& 0  \\
0  &-\hat{h}^*_{BdG}
\end{bmatrix},\nonumber
\end{eqnarray}
where the $3\times 3$ matrix  $\hat{h}_{BdG}$ is defined as
\begin{eqnarray}
\hat{h}_{BdG}=\begin{bmatrix}
\frac{\Gamma}{2}\cos(\phi/2)& i\frac{\Gamma}{2\sqrt{2}}\sin\phi &i\frac{\Gamma}{2\sqrt{2}}\sin\phi   \\
-i\frac{\Gamma}{2\sqrt{2}}\sin\phi  &t-\frac{\Gamma}{4}\cos(\phi/2)& -\frac{\Gamma}{4}\cos(\phi/2)        \\
-i\frac{\Gamma}{2\sqrt{2}}\sin\phi  & -\frac{\Gamma}{4}\cos(\phi/2)  &-t-\frac{\Gamma}{4}\cos(\phi/2)
\end{bmatrix}.\nonumber
\end{eqnarray}
The Green's function $G^{-1}$ can be expressed in terms of Matsubara frequencies $\omega_n=(2n+1)\pi/\beta$ as
\begin{eqnarray}
G^{-1}(\omega_n)=i\omega_n-\hat{H}_{BdG},
\end{eqnarray}
where $\beta=\hbar/(k_b T)$.
The BdG Hamiltonian $\hat{H}_{BdG}$ gives the partition function $Z$ and free energy $F$ as we integrating out the field $\Psi$ and $\Psi^{\dagger}$ in partition function. By defining $\xi_a$ as the eigen energies of $\hat{H}_{BdG}$, we have partition function $Z$ as 
\begin{eqnarray}
Z&=&e^{-\beta F}=\int D\Psi^{\dagger}D\Psi e^{\sum_n \Psi^{\dagger}(i\omega_n-\hat{H}_{BdG})\Psi}\\
&=&\int D\Psi^{\dagger}D\Psi e^{\sum_n\Psi^{\dagger}[G^{-1}(\omega_n)]\Psi}\\
&=&e^{-\sum_n\log \det G^{-1}(\omega_n)}=e^{-\sum_{n,a} \log (i\omega_n-\xi_a)}\\
&=&\Pi_{n,a}(i\omega_n-\xi_a)^{-1}
\label{partitionfunction}
\end{eqnarray}

The free energy $F$ is therefore obtained as 
\begin{eqnarray}
F&=&-\frac{1}{\beta} \ln Z=\frac{1}{\beta}\sum_{n,a} \ln (i\omega_n-\xi_a)
\label{FreeEnergy}
\end{eqnarray}

Summing the Matsubara frequencies gives\cite{altland2010condensed}
\begin{eqnarray}
F&=&-\frac{1}{\beta} \ln Z=\frac{1}{\beta}\sum_{n,a} \ln (i\omega_n-\xi_a)\\
&=&-T\sum^{6}_{a=1} \ln (1+e^{-\beta\xi_a})
\label{FreeEnergy2}
\end{eqnarray}
To get the explicit result of free energy $F$, we evaluate the eigen energy $\xi_a$ of $\hat{H}_{BdG}$ via perturbation theory. By treating the tunneling term $H_{\Gamma}$ as perturbation, the first order perturbation gives the eigen energy\cite{AliceaBraiding} $\xi_a$  as
\begin{eqnarray}
\xi_{M\pm}&=&\pm\frac{\Gamma}{2}\cos(\phi/2)\\
\xi_{p\pm}&=& t\pm\frac{\Gamma}{2}\cos(\phi/2)+\mathcal{O}((\Gamma/t)^2)\\
\xi_{h\pm}&=& -t\pm\frac{\Gamma}{2}\cos(\phi/2)+\mathcal{O}((\Gamma/t)^2)
\label{Eigenenergies}
\end{eqnarray}
Putting the eigen energies showing at Eq.~(\ref{Eigenenergies}) into the free energy showing at Eq.~(\ref{FreeEnergy2}), we get the free energy as a $4\pi$-periodic function. The $\cos(\frac{\phi }{2})$ term in free energy $F$ corresponds to the oscillation of flux $\Phi_B$ of periodicity equal to $2\Phi_0$.  One can check this by using the definition of phase difference of $\phi$ as $\phi=\phi_L-\phi_R=2\pi(\Phi_B/\Phi_0)$. The $\cos(\frac{\phi }{2})$ term has $4\pi$-periodicity in phase difference $\phi$, which corresponds to $2\Phi_0$-periodic in flux $\Phi_B$.

\newpage

\end{document}